\documentclass[preprint,12pt]{elsarticle}

\usepackage{amssymb}
\usepackage{amsmath}
\usepackage{booktabs}
\usepackage{makecell}
\usepackage{multirow}
\usepackage{graphicx}
\usepackage{float}
\usepackage{array}
\usepackage{tabularx}
\usepackage{ragged2e}
\usepackage{placeins}
\usepackage[hidelinks]{hyperref}

\journal{Nuclear Physics B}

\begin{document}

\begin{frontmatter}



\title{Quantitative Analysis of Proxy Tasks for Anomalous Sound Detection}

\author[inst1]{Seunghyeon Shin}
\ead{sh.shin@knu.ac.kr}

\author[inst1,inst2]{Seokjin Lee\corref{cor1}}
\ead{sjlee6@knu.ac.kr}
\cortext[cor1]{Corresponding author.}

\affiliation[inst1]{
    organization={School of Electronic and Electrical Engineering, Kyungpook National University},
    city={Daegu},
    country={Republic of Korea}
}

\affiliation[inst2]{
    organization={School of Electronics Engineering, Kyungpook National University},
    city={Daegu},
    country={Republic of Korea}
}
%

\begin{abstract}
Anomalous sound detection (ASD) often uses proxy objectives or pre-trained encoders because anomalous target-domain samples are scarce. We examine whether proxy-task metrics rank ASD representations across autoencoders, classification, source separation, contrastive learning, and pre-trained models. Frozen representations from nine machine types in the MIMII and ToyADMOS families were evaluated with in-domain and out-domain linear probes and Mahalanobis distance (MD) scoring. Performance was measured by the area under the receiver operating characteristic curve (AUC). No within-family association remained significant after Benjamini--Hochberg correction across 21 exploratory tests. The two smallest unadjusted $p$-values, for autoencoder reconstruction quality and ArcFace macro-$F_1$ with MD AUC, both yielded $q=0.181$. The evaluated metrics showed rank reversals. The model with the highest scale-invariant signal-to-distortion ratio (SI-SDR) was not best under any ASD protocol. The lowest reconstruction error and the highest macro-$F_1$ or AudioSet mean average precision (mAP) did not consistently identify the best representation. Cross-entropy classification nevertheless transferred well, achieving the highest in-domain, out-domain, and MD AUCs of 75.15\%, 67.36\%, and 67.16\% under the shared EfficientNet-Lite encoder family at different capacities. This comparison standardizes the encoder family but not the task pipeline, because the shared separation branch uses a mel-domain reconstruction adapter rather than the task-specific short-time Fourier transform (STFT) separator. The change in separation transfer therefore cannot be attributed to the encoder alone. Proxy-best checkpoint selection did not systematically outperform minimum-loss selection. The results indicate that proxy-metric ranking, objective transfer, and task-interface dependence should be evaluated separately.

\end{abstract}




\end{frontmatter}



\section{Introduction}
\label{sec1}

Acoustic-based condition monitoring has been adopted across various domains as microphone sensors are cost-effective and can operate in non-contact, occluded, or unlit environments. Acoustic monitoring can be categorized into two main streams, namely, sound event detection (SED)  for surveillance~\cite{valenzise2007scream, clavel2008detection} and anomalous sound detection (ASD) for monitoring the continuous operating noise generated from machinery~\cite{koizumi2018unsupervised, jardine2006review}. Systems used for ASD are typically trained using only normal operating sounds, addressing the scarcity of anomalous data and the challenges associated with detecting unseen anomalous states~\cite{koizumi2020description}. Therefore, the system learns the distribution of normal sounds and determines an anomalous state by quantifying the deviation of the input signal from the normal distribution to detect anomalies using only normal data~\cite{kawachi2018complementary}.

Neural networks must be trained exclusively on normal state audio; therefore the ultimate objective of detecting anomalies cannot be learned directly through supervised approaches. This fundamental constraint necessitates the use of proxy tasks---auxiliary objectives that can be optimized using only normal data while implicitly learning representations useful for distinguishing anomalies. Consequently, proxy tasks are employed to indirectly learn representations to detect anomalies. Contemporary ASD research employs diverse proxy tasks. These include AutoEncoders (AE) that learn to reconstruct input signals~\cite{marchi2015nonlinear}, classification that leverages machine IDs or operating conditions~\cite{Giri2020a}, contrastive learning that acquires augmentation-invariant representations~\cite{hojj2022cont}, source separation that isolates target sounds from mixtures~\cite{lee2024activity}, and transfer learning that employs pre-trained models~\cite{ANVAR2023103872}. These methods operate on distinct learning principles, namely, reconstruction, discrimination, and invariance learning; however, these principles share a common goal of capturing normal sound characteristics.

Despite this methodological diversity, the relationship between proxy task performance and ASD capability has received limited attention. For example, in the case of the widely adopted AE approach, the premise that anomalies increase reconstruction errors is consistently stated~\cite{marchi2015novel}; however, the extended hypothesis that optimizing reconstruction quality improves detection performance remains empirically unverified. Similarly, whether tight feature clustering from metric learning (e.g., ArcFace~\cite{Deng_2022}) improves anomaly detection has not been tested. Moreover, researchers have observed that pretext task performance does not always predict downstream success in computer vision~\cite{kolesnikov2019revisiting}. Whether similar patterns apply to ASD has yet to be systematically examined.

Understanding this relationship carries significant practical implications. Specifically, if proxy task optimization fails to improve ASD, then the common practice of designing and tuning proxy tasks based solely on their intrinsic metrics may be fundamentally misguided. Conversely, proxy metrics could serve as reliable surrogates for ASD performance during model development in the presence of strong correlations, reducing the reliance on anomalous samples in the validation phase. The relevant methodological question is whether a proxy task optimized using only normal data yields representations aligned with the downstream ASD objective. We attempt to answer this question by testing whether rankings induced by proxy-task metrics agree with those based on downstream ASD performance across systematically varied model configurations evaluated under a common set of protocols. This requires separating three issues that ASD studies have typically treated together: whether an intrinsic proxy metric ranks model variants as ASD performance does, whether the proxy objective yields representations that transfer to ASD regardless of that metric, and whether such conclusions persist when the encoder and task interface change.
This study makes three distinct contributions: First, we test whether intrinsic proxy-task metrics rank ASD representations consistently with downstream performance across five families: AE, classification, source separation, contrastive learning, and pre-trained models. We compare these metrics through linear-probe (LP) and Mahalanobis distance (MD) evaluations across nine machine types from the ToyADMOS and MIMII dataset families, comprising the seven machine types in the DCASE 2022 configuration, along with Pump and ToyConveyor in their DCASE 2020 configurations. Second, as an additional analysis, we retrain four proxy-task families from scratch with a common EfficientNet-Lite encoder family, thereby reducing, although not eliminating, between-task architectural variation. Third, we distinguish proxy-metric ranking from objective transfer and task-interface dependence and propose a three-stage procedure that operationalizes these distinctions, covering proxy-task health and metric resolution, objective transfer to ASD, and ranking and interface robustness.

The remainder of this paper is organized as follows: Section~\ref{sec2} reviews ASD and proxy--target alignment. Section~\ref{sec3} describes the proxy tasks and experimental configurations. Section~\ref{sec4} presents the evaluation protocols. Section~\ref{sec5} reports the task-specific, pre-trained, shared-backbone, and checkpoint-sensitivity results. Section~\ref{sec6} discusses proxy-metric ranking and objective transfer, along with the limitations and practical implications of this study. Finally, Section~\ref{sec7} concludes the paper.

\section{Related Work}
\label{sec2}
\subsection{General Unsupervised Anomalous Sound Detection}

In the early stages, acoustic monitoring was primarily employed in surveillance applications to detect specific sound events such as gunshots, screams, or car crashes~\cite{valenzise2007scream, clavel2005events, foggia2015audio}. In these scenarios, collecting anomalous samples is relatively straightforward. In contrast, condition monitoring, to detect deviations from a normal state, was previously limited to specialized tasks such as heart sound detection~\cite{rubin2017recognizing} or monitoring specific machinery~\cite{Lee2016, Park2018} due to the difficulty in acquiring anomalous data.

However, the release of public datasets such as ToyADMOS~\cite{koizumi2019toyadmos} and MIMII~\cite{purohit2019mimii}, and the establishment of the Detection and Classification of Acoustic Scenes and Events (DCASE) challenges~\cite{koizumi2020description}, have considerably increased the volume of research in this field. These datasets typically comprise normal operating sounds for training, whereas anomalous sounds, often recorded by intentionally damaging the machinery, are provided strictly for performance evaluation. To facilitate the development of highly robust systems suitable for real-world applications, ToyADMOS2~\cite{harada2021toyadmos2} was subsequently released to evaluate performance under domain shifts. A key design characteristic of these public datasets is the provision of several normal operating sounds for training, enforcing an unsupervised learning paradigm in which anomalous samples are reserved exclusively for the testing phase.

Recent ASD benchmarks also include first-shot generalization to unseen machine types, domain shifts, and environmental noise~\cite{Nishida2025DCASE,Nishida2026DCASE}. These settings test whether proxy metrics optimized on normal training data remain associated with ASD representation quality under distribution shifts.

\subsection{Proxy Tasks for Unsupervised Anomalous Sound Detection}

In ASD environments restricted to normal operating sounds, direct training for anomaly detection is infeasible. Researchers employ various proxy tasks such that neural networks can learn the inherent characteristics of normal data, ultimately yielding features that distinguish anomalous inputs from learned normal patterns.

Reconstruction-based methods, particularly AEs, have been widely adopted since the establishment of the DCASE challenge baseline~\cite{Dohi2022-2}. These methods assume that models trained on normal data will produce increased reconstruction errors for anomalies~\cite{guan2023transformer, kuroyanagi2021anomalous}.
Source separation-based approaches train neural networks to isolate target machine sounds from mixtures containing interfering noise. Recent studies on source separation-based ASD have demonstrated that separation models trained on normal sounds can be leveraged to detect anomalies~\cite{Shin2024, lee2024activity}.

Discriminative approaches leverage auxiliary information such as machine IDs and operating conditions~\cite{wu2023unsupervised}. Recent studies on metric learning for ASD have demonstrated that metric learning objectives such as ArcFace~\cite{Deng_2022} enhance feature compactness, thereby improving detection performance~\cite{wilkinghoff2023angular}. Recent discriminative methods supplement fixed machine-ID supervision with metadata-derived attributes learned through masked reconstruction, self-distillation, and pseudo-label prediction~\cite{Jiang2025GLAMASD}.

Self-supervised methods, including contrastive learning frameworks such as SimCLR~\cite{chen2020simple} and SimSiam~\cite{chen2021exploring}, aim to learn augmentation-invariant representations~\cite{hojj2022cont, guan2023anomalous}. Conversely, transfer learning approaches employ models pre-trained on large-scale datasets. Specifically, models such as BEATs~\cite{chen2022beats} and EAT~\cite{chen2024eat} have demonstrated strong performances as feature extractors in recent ASD studies~\cite{WangMYPS2025, SaengthongSCITOK2025}. Recent work has also adapted large pre-trained audio encoders to generalized ASD through parameter-efficient modules and contrastive objectives, while training-free studies show that pooling and feature aggregation can materially affect downstream performance \cite{Han2025GeneralizedSSLASD,Wilkinghoff2026TemporalPooling}. In addition, audio tagging models such as CED~\cite{Dinkel2024CED} provide multiple pre-trained variants with different AudioSet mAP (mean average precision) values, making them useful for examining whether general audio classification performance predicts ASD transferability.

Table~\ref{tab:related_work_summary} summarizes the assumptions underlying these proxy tasks. Prior studies have usually reported final ASD performance rather than the relation between intrinsic proxy metrics and downstream performance. Section~\ref{sec3} describes the implementations and experimental design used in this study.

\subsection{Evaluating Representations in Unsupervised Learning}

Representations learned by neural networks are high-dimensional; therefore, directly evaluating their quality for ASD is inherently difficult. Consequently, indirect evaluation methods are employed to compare these representations, broadly categorized into qualitative and quantitative approaches.

Qualitative evaluation is typically conducted using visualization techniques based on dimensionality reduction, such as t-SNE~\cite{maaten2008visualizing} or UMAP~\cite{mcinnes2018umap}. These methods project the high-dimensional feature space onto a low-dimensional plane, allowing the clustering of compact data points of the same class and the separation from other classes to be visually inspected. Although these visualizations offer intuitive insights, they are sensitive to hyperparameter settings and do not yield numerical metrics, complicating the direct quantitative comparisons between different systems.

Quantitative evaluation involves attaching an auxiliary module to the learned representations to derive numerical metrics. The most widely adopted approach is the LP~\cite{alain2016understanding}, which assesses the quality of features by training a single linear classifier on top of the fixed (frozen) output features of the pre-trained network, utilizing a labeled evaluation dataset. The resulting performance indicates the ease with which the representations can be separated by a linear decision boundary.

Although high linear separability is generally indicative of a well-trained network, the quality of a representation does not necessarily guarantee linear separability~\cite{hewitt2019designing}. Learned representations are often evaluated using alignment and uniformity metrics in the domain of unsupervised learning, particularly contrastive learning~\cite{wang2020understanding}. Alignment measures the proximity of features from augmented views of the same sample, whereas uniformity assesses the even distribution of features from different samples. However, these metrics are intrinsically designed for frameworks that depend heavily on data augmentation. Consequently, they are ill-suited for evaluating other proxy tasks, such as those based on reconstruction-based objectives (e.g., AE and source separation), in which such augmentation strategies are not central to the learning process. 
Clear metric reporting is also emphasized in related biomedical signal-classification studies, such as automatic heartbeat classification, where the evaluation protocol and classification metrics are described explicitly to support methodological transparency~\cite{heartbeat}.
ASDKit provides a unified framework for heterogeneous ASD methods, supporting consistent representation-level evaluation~\cite{Fujimura2025ASDKit}.

\subsection{Proxy-Target Alignment in Other Domains}

The relationship between proxy (or pretext) and downstream task performances has been investigated in the domain of machine learning, particularly in computer vision. A large-scale evaluation~\cite{ericsson2021well} of 13 self-supervised models across 40 downstream tasks revealed that the accuracy of ImageNet Top-1 correlates well with many-shot recognition but poorly with few-shot learning, object detection, and dense prediction tasks. This study demonstrated that high pretext task performance does not universally guarantee downstream success. Similarly, research addressing the specificity of proxy tasks has demonstrated that features optimized for pretext tasks may become overly specialized, limiting the generalization of these features to target tasks~\cite{kolesnikov2019revisiting}. These findings from the domain of computer vision suggest that the assumed link between proxy and target task performance warrants careful examination.
In the ASD domain, however, such systematic analysis has not been conducted. Although the foundational assumption of reconstruction-based methods, i.e., anomalies increase reconstruction errors, is explicitly stated in DCASE baseline system descriptions~\cite{koizumi2020description}, the extended hypothesis, i.e., optimizing reconstruction quality improves ASD performance, has not been empirically tested. To address this gap, we analyze five proxy-task families, the assumptions of which are summarized in Table~\ref{tab:related_work_summary}, along with the corresponding unresolved questions.

\begin{table}[t]
\centering
\caption{Summary of representative proxy-task research streams, their underlying
proxy assumptions, and the remaining gaps addressed in this study.}
\label{tab:related_work_summary}
\footnotesize
\setlength{\tabcolsep}{4pt}
\renewcommand{\arraystretch}{1.08}
\begin{tabular}{@{}>{\RaggedRight\arraybackslash}p{0.24\linewidth}
                >{\RaggedRight\arraybackslash}p{0.70\linewidth}@{}}
\toprule
\textbf{Research stream} & \textbf{Proxy assumption and remaining gap} \\
\midrule
\textbf{Reconstruction}\par{\scriptsize \cite{Dohi2022-2,kuroyanagi2021anomalous,guan2023transformer}}
& Models trained on normal data are assumed to produce larger reconstruction errors for anomalies; however, whether improved normal reconstruction serves as a reliable surrogate for ASD capability has not been verified. \\
\addlinespace[0.35em]
\textbf{Discriminative / metric learning}\par{\scriptsize \cite{Giri2020a,wu2023unsupervised,wilkinghoff2023angular,Jiang2025GLAMASD}}
& Auxiliary labels, angular margins, or metadata-derived attributes encourage compact or structured representations, yet saturated auxiliary scores may not reflect ASD-oriented representation quality. \\
\addlinespace[0.35em]
\textbf{Contrastive learning}\par{\scriptsize \cite{hojj2022cont,guan2023anomalous,Han2025GeneralizedSSLASD}}
& Augmented views or contrastive adaptation objectives are used to learn invariant acoustic representations, but the relationship among contrastive metrics, representation collapse, and ASD performance remains unclear. \\
\addlinespace[0.35em]
\textbf{Source separation}\par{\scriptsize \cite{Shin2024,lee2024activity}}
& Target machine sounds are separated from interference to model normal acoustic structure; whether separation quality quantitatively tracks ASD capability remains insufficiently examined. \\
\addlinespace[0.35em]
\textbf{Pre-trained audio models}\par{\scriptsize \cite{ANVAR2023103872,chen2022beats,chen2024eat,Dinkel2024CED,WangMYPS2025,SaengthongSCITOK2025,Han2025GeneralizedSSLASD,Wilkinghoff2026TemporalPooling}}
& Large-scale audio representations are transferred to ASD, yet general audio benchmark scores, adaptation strategies, and pooling choices may not directly predict ASD transferability. \\
\addlinespace[0.35em]
\textbf{Representation evaluation / alignment}\par{\scriptsize \cite{alain2016understanding,lee2018simple,wang2020understanding,kolesnikov2019revisiting,ericsson2021well}}
& Linear probing, distance-based scoring, and transfer analyses assess representation quality, but they have rarely been integrated to verify proxy--ASD alignment across proxy task families. \\
\bottomrule
\end{tabular}
\end{table}

\section{Proxy Tasks and Configurations}
\label{sec3}
\subsection{Overview}
\label{sec31}

To systematically investigate the relationship between proxy task performance and ASD capability, we selected five proxy task families that represent the diverse learning paradigms currently dominant in ASD research.
Table~\ref{tab:proxy_task_summary} summarizes the input data, training objective, and ASD representation for each proxy task. The families differ in backbone capacity, feature dimension, and extraction method, which hinders direct comparison. We therefore conduct a shared-backbone analysis (Section~\ref{sec:shared_backbone}) with AE, source separation, classification, and contrastive learning using a common EfficientNet-Lite~\cite{tan2020efficientnet} encoder family. This analysis controls the encoder family while retaining task-specific heads and output formulations.

First, we employ AE and source separation, representing the reconstruction-based paradigm. These methods rely on the hypothesis that models trained on normal data will yield large errors or degrade outputs when processing anomalies, thereby manifesting distinguishable differences in the feature space.
Second, we employ classification tasks that leverage auxiliary attribute information, such as machine IDs, for the discriminative paradigm. Recent approaches in this domain frequently incorporate metric learning objectives, such as ArcFace~\cite{Deng_2022}, to enforce compact intra-class clustering and enhance decision boundaries.
Third, we adopt contrastive learning frameworks, specifically SimCLR~\cite{chen2020simple} and SimSiam~\cite{chen2021exploring}, representing the self-supervised paradigm. These methods focus on learning invariant representations by maximizing agreement between augmented views of the same sample, essential for evaluating feature robustness in the absence of explicit labels.
Finally, to address the transfer learning paradigm, we evaluate pre-trained models. We employ representations from large-scale foundational models to assess the transferability of general acoustic features to the specific domain of ASD.

Unless stated otherwise, models trained from scratch use the same optimizer and are trained for the same duration. All neural networks are trained using the AdamW optimizer~\cite{loshchilov2017decoupled}, coupled with a StepLR scheduler for 200 epochs. We select models based on the lowest training loss rather than employing a separate validation set to maximize the use of limited training data for learning ASD-suitable representations and strictly prevent information leakage from anomalous samples during the validation phase. Unless otherwise noted in the respective subsections, input signals are processed as 10-s audio clips (16\,kHz) converted to log-mel spectrograms with 128 mel bins, using an FFT (fast Fourier transform) size of 1024 and a hop size of 512.

\begin{table*}[t]
\centering
\caption{Overview of the task-specific proxy task implementations used in this study. The 'Learning objective (loss)' column specifies the training loss, which does not necessarily coincide with the proxy evaluation metric (Section~\ref{sec43}). The 'Feature for ASD' column specifies the representation extracted for anomaly detection.}
\label{tab:proxy_task_summary}
\resizebox{\textwidth}{!}{%
\begin{tabular}{l l l l l}
\toprule
\textbf{Learning paradigm} & \textbf{Proxy task} & \textbf{Input data} & \textbf{Learning objective (loss)} & \textbf{Feature for ASD (output)} \\
\midrule
\multirow{2}{*}{Reconstruction} & AE & Log-mel Spectrogram & Mean squared error & Element-wise absolute reconstruction error \\
 & Source separation & Mixture Spectrogram & Spectral mean squared error & Concatenated channel-pooled features \\
\midrule
Discriminative & Classification & Log-mel spectrogram & Cross-entropy / ArcFace & Backbone output (penultimate layer) \\
\midrule
Self-supervised & \begin{tabular}[c]{@{}l@{}}Contrastive learning\\(SimCLR, SimSiam)\end{tabular} & Augmented views & \begin{tabular}[c]{@{}l@{}}NT-Xent (SimCLR)\\Neg. Cosine Sim. (SimSiam)\end{tabular} & Backbone output (ResNet) \\
\midrule
Transfer learning & Pre-trained models & Model-specific audio frontend & (Pre-trained on AudioSet) & Mean-pooled embeddings \\
\bottomrule
\end{tabular}%
}
\end{table*}
\FloatBarrier
\subsection{AutoEncoder}
\label{sec32}
AEs have served as the foundational baseline for the DCASE Challenge Task 2~\cite{Dohi2022-2} series since 2020, establishing the standard benchmark for unsupervised ASD. Including AEs is, therefore, essential to provide a reference point for performance comparison in our analysis. Although simple, reconstruction-based methods remain effective, relying on the premise that models trained to minimize reconstruction error on normal data will increase errors under anomalous inputs that deviate from the learned distribution. The underlying hypothesis is that improved reconstruction of normal patterns should clarify the separation from anomalies, thereby improving ASD performance.

The fundamental objective of an AE is to learn a compressed representation of the input signal and subsequently reconstruct the original signal from this latent code. During training, the network optimizes its parameters to minimize the discrepancy between the input and the reconstructed output. We minimize the mean squared error (MSE) between the input and reconstructed log-mel spectrograms. Reconstruction quality is evaluated based on the mean absolute error (MAE), which is defined in Section~\ref{sec431}. We adopt a fully connected (dense) AE architecture consistent with the DCASE 2022 Task 2 baseline~\cite{Dohi2022-2}. The encoder and decoder each consist of five fully connected layers. The input is a 640-dimensional vector formed by concatenating five consecutive frames of a 128-bin log-mel spectrogram. Moreover, the encoder progressively compresses this input through hidden layers to a bottleneck latent space, and the decoder reconstructs the original representation from this latent code. This bottleneck constraint compels the network to prioritize the dominant patterns of the normal training sound distribution, thereby inhibiting the accurate reconstruction of unseen irregular patterns.

Unlike typical representation learning, which employs the latent vector, reconstruction-based ASD employs the reconstruction error as the feature representation, following the DCASE baseline system configuration. We use the element-wise absolute reconstruction error vector, which captures per-bin discrepancies between the input and output spectrograms, directly reflecting the inability of the model to reconstruct anomalous time-frequency patterns.

To systematically analyze the effect of model capacity on the proxy-ASD relationship, we vary the hidden layer (64, 128, 256) and bottleneck (4, 8, 16) dimensions, yielding nine experimental configurations with different compression ratios.

\subsection{Classification}
 \label{sec33}
Classification is a dominant proxy task in the ASD domain. Prior studies have demonstrated that incorporating granular attribute information, such as operation speeds, machine loads, or domain labels, can enhance detection performance~\cite{wilkinghoff2023angular, wu2023unsupervised}. However, in this study, we restrict the scope to standard machine ID classification, excluding detailed attribute metadata, to evaluate the fundamental discriminative capability derived solely from acoustic signals and to ensure a generalizable assessment across different deployment scenarios. The hypothesis is that tight intra-class clustering in the feature space should yield highly discriminative representations for anomaly detection.

We employ two representative training objectives. The first is the standard cross-entropy loss, which trains the network to classify inputs by maximizing the predicted probability of the ground-truth class. The second is ArcFace (additive angular margin loss)~\cite{Deng_2022}, a metric learning objective that imposes an additive angular margin penalty. This explicitly enforces intraclass compactness within the feature space, which is expected to improve the distinguishing of anomalies from normal patterns.

We employ the ResNet family (ResNet-18, 34, 50, 101, and 152)~\cite{he2016deep}, a standard choice for extracting deep acoustic representations, for the backbone architecture. Within the ResNet family, ResNet-18 and ResNet-34 use basic residual blocks and output 512-dimensional backbone features, whereas ResNet-50, ResNet-101, and ResNet-152 use bottleneck residual blocks and output 2048-dimensional backbone features. The structural difference between the two objectives is related to the auxiliary head attached to the backbone. Specifically, CE uses standard features for linear layer mapping to class logits, whereas ArcFace employs a margin-based projection layer to compute angular distances. These auxiliary heads are used exclusively during proxy task training. For ASD evaluation, we discard these heads and employ the feature vector from the penultimate layer (the ResNet backbone output) as the representation, yielding feature dimensions of 512 for ResNet-18/34, and 2048 for ResNet-50/101/152.
The combination of five backbone variants and two loss functions yields ten experimental configurations, enabling the effects of network capacity and training objective on the proxy-ASD relationship to be analyzed. 

\subsection{Source Separation}

Source separation represents a distinct paradigm in unsupervised ASD, diverging from standard compression or discrimination approaches. Preliminary studies have investigated the utility of source separation for ASD; however, the quantitative relationship between separation quality and detection performance has not yet been systematically analyzed~\cite{shimonishi2023anomalous, Shin2024}. We include this task herein to investigate whether the objective of isolating target sounds, distinct from reconstruction or classification, yields representations beneficial for ASD. We posit that improved separation capability correlates with a highly precise characterization of normal sound patterns, thereby enhancing the detection of deviations.

The core objective is to estimate the target machine's clean signal from a synthetic mixture. During training, the model receives a mixture of the target normal sound and randomly sampled auxiliary noise at varying signal-to-noise ratios (-5--5 dB). By learning to filter out these variable noise components, the network implicitly captures the inherent structural patterns of normal sounds required to accurately reconstruct signals. The network is trained on the spectral MSE between the estimated and clean target spectrograms. Separation performance is evaluated in terms of SI-SDR, which is defined in Section~\ref{sec433}.

We adopt the architecture proposed in~\cite{Shin2024}, based on the CMGAN framework~\cite{CMGAN}. This architecture features a Dilated DenseNet-based encoder and decoder, with conformer~\cite{gulati2020conformer} blocks interposed between them to enhance temporal modeling.
Unlike the AE, which directly employs the reconstruction error, we leverage the intermediate representations learned by the separation network. Specifically, we extract feature maps from the output of the dense encoder and each subsequent conformer block; for a model with $N$ conformer blocks, features are aggregated from $N+1$ extraction points. Due to the computational complexity of processing full 10-s sequences, the network operates on 2-s segments. Channel-wise average pooling is applied to each extraction point, and the final representation is obtained by concatenating these pooled vectors along the time axis, capturing multi-scale structural characteristics.
To analyze the relation of separation capability with ASD performance, we vary the number of conformer blocks (0, 1, 2, 4) and the channel width (64, 128), yielding eight experimental configurations.

\subsection{Contrastive Learning}
\label{35}

Contrastive learning is a widely adopted self-supervised methodology that leverages data augmentation to learn representations without explicit labels. We include this task to systematically investigate whether augmentation-invariant representations benefit anomaly detection. The hypothesis is that learning to capture intrinsic acoustic patterns invariant to augmentation may distinguish normal sounds from anomalies.

The fundamental objective is to learn consistent representations from the same input regardless of applied augmentations. To generate augmented views, we apply a combination of time-domain masking, frequency-domain masking, pitch shifting, time stretching, and white noise injection. These augmentations are selected to cover both spectral and temporal variations commonly encountered in acoustic signals. We employ two representative frameworks. SimCLR~\cite{chen2020simple} maximizes the similarity between positive pairs (augmented views from the same sample) while minimizing the similarity between negative pairs (views from different samples), using NT-Xent loss. In contrast, SimSiam~\cite{chen2021exploring} relies solely on positive pairs without requiring negative samples.

For the backbone architecture, we employ the ResNet family (ResNet-18, 34, 50, 101, and 152)~\cite{he2016deep}. The same ResNet family and feature-dimension split described in Section~\ref{sec33} are used for contrastive learning. An MLP projection head is attached to the backbone during training, consisting of two hidden layers (Linear--BatchNorm--ReLU) followed by a final linear layer that projects features into a 128-dimensional space where the contrastive objective is applied. This projection head is discarded for ASD evaluation; the feature representation is extracted from the backbone output (penultimate layer), yielding 512-dimensional vectors for ResNet-18/34 and 2048-dimensional vectors for ResNet-50/101/152.
The combination of five backbone variants and two contrastive frameworks yields ten experimental configurations, enabling analysis of the effect of model capacity and learning strategy on the proxy-ASD relationship.

\subsection{Shared-Backbone Configuration}
\label{sec:shared_backbone}

\begin{figure}
    \centering
    \includegraphics[width=1.0\linewidth]{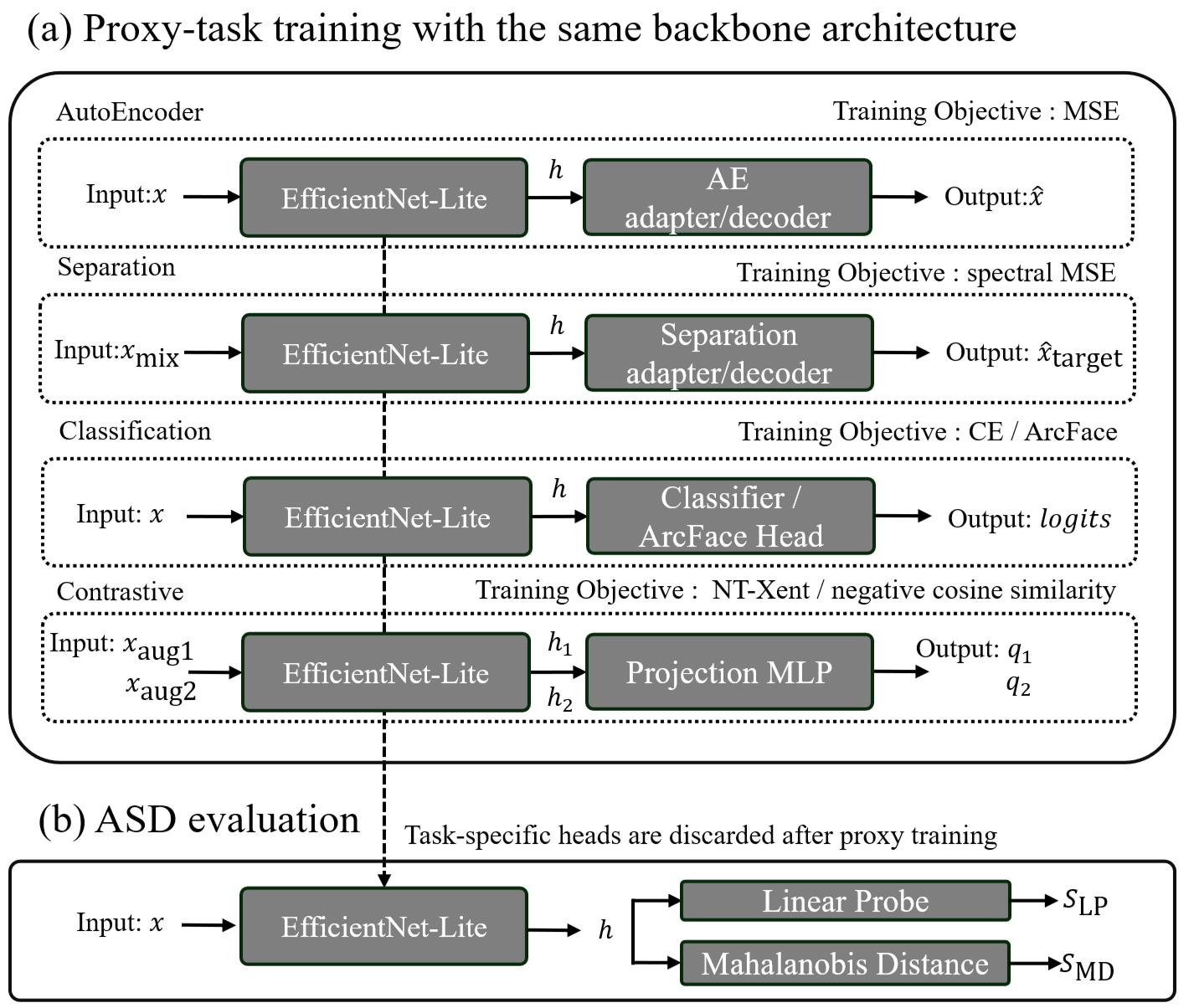}
    \caption{Shared-backbone configuration. (a) AE, mel-domain component reconstruction, classification, and contrastive learning are trained using EfficientNet-Lite encoders and task-specific heads. (b) The heads are removed after proxy training, and the backbone embedding $h=F(x)$ is evaluated via LP and MD scoring. This design standardizes the encoder family while retaining task-specific adapters and outputs.
}
    \label{fig:0}
\end{figure}

The task-specific implementations in the primary analysis involve a fully connected AE, ResNet backbones, and a CMGAN-based separation network. As part of  an additional analysis, we retrain AE, source separation, classification, and contrastive learning from scratch with EfficientNet-Lite0--Lite4 encoders~\cite{tan2020efficientnet} (Fig.~\ref{fig:0}). This design reduces between-task architectural variation and tests whether the proxy--ASD ranking patterns persist within a common encoder family. Because task-specific heads, adapters, and output formulations are retained, this experiment reduces architectural variation but does not fully isolate the effect of the proxy objective.

The common backbone maps a log-mel spectrogram $x$ to an embedding $h=F(x)$. Each task includes a separate training head. The AE head reconstructs the flattened log-mel input with a DCASE baseline-style decoder~\cite{Dohi2022-2}. The separation head reconstructs element-wise mel-domain components from the mixture representation. Both heads are trained on MSE; their proxy metrics are the input-reconstruction MAE for AE and the component-reconstruction MAE for separation. The shared separation metric therefore differs from the SI-SDR used by the task-specific STFT-domain separator. SI-SDR is a time-domain, waveform-level metric and is therefore not defined for this branch, which reconstructs mel-domain components rather than a time-domain waveform; consequently, we evaluate it based on the component-reconstruction MAE. Classification and contrastive learning respectively use a linear or ArcFace head and the projection head described in Section~\ref{35}. The shared separation branch is a mel-domain component-reconstruction adapter rather than a direct reimplementation of the task-specific separator.

After proxy training, the task-specific heads are removed and the backbone embedding $h$ is evaluated via LP and MD scoring. All shared-backbone tasks use the same feature pathway. This setting differs from the task-specific AE, which uses an element-wise absolute reconstruction error vector as its ASD representation. The encoder capacity ranges from Lite0 to Lite4. This design standardizes the encoder family but not the complete task pipeline.

\FloatBarrier
\subsection{Pre-trained Models}

Pre-trained models have demonstrated strong performances in recent ASD studies, serving as feature extractors or teacher models for knowledge distillation. We include this paradigm to examine whether general audio understanding, as measured by AudioSet~\cite{gemmeke2017audio} classification performance, transfers effectively to ASD. The implicit assumption in adopting pre-trained models is that increased classification accuracy on large-scale audio datasets should yield increased discriminative representations for anomaly detection.
Unlike other proxy tasks trained from scratch, we evaluate off-the-shelf models without additional fine-tuning on the target dataset. This allows us to assess the intrinsic transferability of pre-trained representations to the ASD domain.

We evaluate three pre-trained encoder families: BEATs~\cite{chen2022beats}, EAT~\cite{chen2024eat}, and CED~\cite{Dinkel2024CED}. BEATs uses iterative audio pre-training with acoustic tokenizers, EAT uses masked audio modeling, and CED uses consistent ensemble distillation for AudioSet tagging. While these families differ in their pre-training objective, frontend, and capacity, they all report AudioSet AS-2M mAP as a source-task reference.

We evaluate BEATs-iter3, BEATs-iter3+, EAT-base, EAT-large, CED-tiny, CED-mini, CED-small, and CED-base. No encoder is fine-tuned on the target dataset, and temporal or model-specific pooling yields one representation per clip. Temporal mean pooling produces 768-dimensional representations for BEATs and EAT-base and a 1536-dimensional representation for EAT-large. The CED extraction settings and output dimensions match the released configurations and are reported in the accompanying repository.

\section{Evaluation Methodology}
\label{sec4}
\subsection{Evaluation Protocol}

To evaluate the representations consistently learned by feature extractor $F$, we employ a unified evaluation protocol. Assessing these representations involves two essential steps as follows: deriving an anomaly score from the extracted feature vectors and quantitatively evaluating the overall detection performance of the system based on these calculated scores.

We adopt simple scoring mechanisms that avoid intricate post-processing or reliance on auxiliary information to ensure that the evaluation focuses precisely on the quality of the learned representations, rather than on the complexity of the scoring backend. We employ two distinct, complementary scoring protocols for feature evaluation: LP, which utilizes a single linear classifier to assess the linear separability of the representations, and MD~\cite{lee2018simple}, which evaluates the distributional compactness of the normal features. We quantify the final ASD performance using the area under the receiver operating characteristic curve (AUC). AUC is widely adopted in ASD research as it provides a robust, threshold-independent measure of the system's ability to distinguish between normal and anomalous samples.

\begin{figure}[H]
    \centering
    \includegraphics[width=0.95\linewidth]{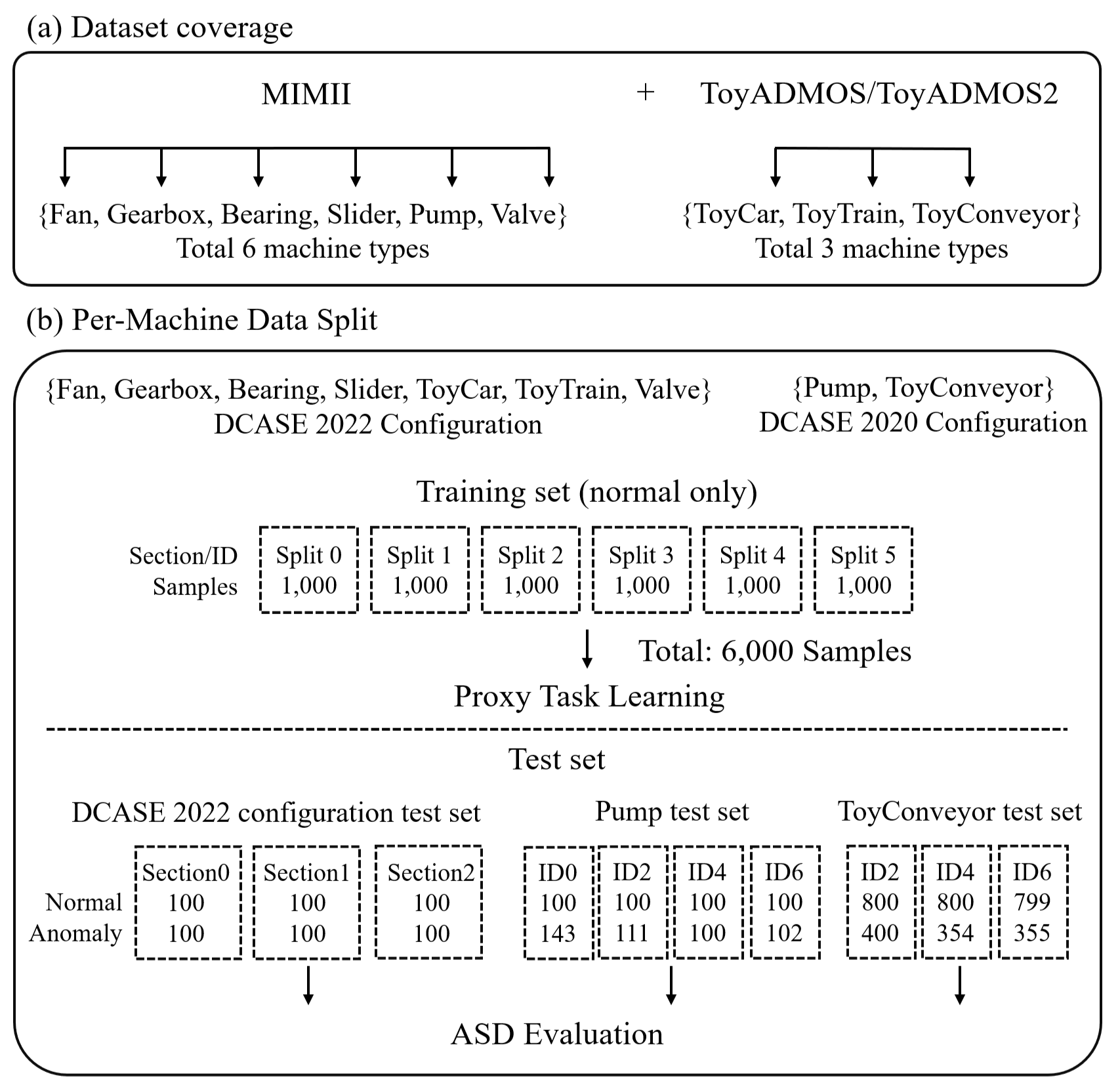}
    \caption{Dataset coverage and per-machine data splits used in this study. (A) Nine machine types drawn from the MIMII and ToyADMOS dataset families. (B) The seven machine types of the DCASE 2022 configuration use section-based splits (training: six sections $\times$ 1,000 normal samples; evaluation: three sections, each with 100 normal and 100 anomalous samples). Pump and ToyConveyor follow their DCASE 2020 configurations, in which machine IDs are treated as sections; the corresponding per-ID evaluation counts are shown.}
    \label{fig:dataset_config}
\end{figure}

We performed our evaluation using the MIMII~\cite{purohit2019mimii} and ToyADMOS~\cite{koizumi2019toyadmos}/ToyADMOS2~\cite{harada2021toyadmos2} dataset families (Fig.~\ref{fig:dataset_config}). Seven machine types—fan, gearbox, bearing, slider, ToyCar, ToyTrain, and valve—were assessed, in line with the DCASE 2022 Task 2 configuration. A section denotes a subset associated with a domain-shift condition, such as machine parameters, acoustic environment, post-maintenance state, or recording device. For Pump and ToyConveyor, we employ their DCASE 2020 configurations, in which subsets are organized by machine ID; each ID is treated as a section for these two machine types.

The LP scenarios differ in whether classifier training and evaluation use the same section domains. In-domain evaluation uses disjoint samples from the same sections, whereas out-domain evaluation holds out complete sections from classifier training.  For the seven DCASE 2022 machine types, the proxy-training set contains 6,000 normal samples from six sections. The LP evaluation set contains 600 samples from three sections, with 100 normal and 100 anomalous samples per section. Pump and ToyConveyor retain the native per-ID counts of their DCASE 2020 configurations as shown in Fig.~\ref{fig:dataset_config}(b).

All experiments were implemented in PyTorch 2.7.1 and run on a workstation equipped with an NVIDIA GeForce RTX 5090 GPU. The source code, configuration files, model-specific hyperparameters, and runtime records are available at \url{https://anonymous.4open.science/r/asd-proxy-task-analysis-3628/}.

To quantitatively analyze the correlation between proxy task performance and ASD performance, we employ the non-parametric Spearman's rank correlation coefficient. Given the limited sample size, $p$-values were computed using an exact test to ensure statistical validity.

Within each task-specific family, we compute the Spearman rank correlations between the full-precision proxy metric and each downstream ASD metric. MAE and uniformity are sign-reversed to ensure that larger values indicate higher proxy quality. Exact permutation $p$-values are reported because each family contains five to nine configurations. The configurations are designed architecture variants rather than independent random samples. The coefficients and $p$-values are therefore treated as descriptive. Shared-backbone configurations are analyzed separately. The 21 tests are treated as a single multiplicity family because each evaluates whether a proxy metric ranks downstream ASD performance. The pre-trained group is interpreted separately because its mAP values are externally reported and because its encoders differ in their pre-training objective, architecture, pooling, and feature dimension.

\subsection{Anomaly Score Calculation}
\subsubsection{Linear Probe}
LP is a widely adopted technique to evaluate representation quality by measuring the linear separability of features extracted from a trained network. In this protocol, a single linear layer is appended to the pre-trained feature extractor, the weights of which remain frozen. This linear layer is subsequently trained to perform a binary classification task, distinguishing between normal and anomalous states. For input acoustic signal $x$, feature representation $h$ is extracted as defined in Section~\ref{sec3}. Consequently, $h$ corresponds to the reconstruction error of the AE, the penultimate layer output for classification and contrastive learning, or the concatenated features for source separation. In the shared-backbone configuration, $h$ consistently denotes the EfficientNet-Lite backbone embedding for all proxy tasks, as described in Section~\ref{sec:shared_backbone}. Formally, we define the feature vector as $h=F(x)$. Subsequently, this vector is passed through the linear classifier to obtain logits $\mathbf{z}$, as defined below.

\begin{equation}
    \label{eq1}
    \mathbf{z} = \mathbf{W}\mathbf{h} + \mathbf{b},
\end{equation}
where $\mathbf{W}$ and $\mathbf{b}$ denote the weight matrix and the bias vector of the linear layer, respectively. These parameters are updated during the classifier training phase. Training is conducted using a subset of the evaluation data containing both normal and anomalous samples, to minimize the cross-entropy loss function defined as

\begin{equation}
    \label{eq2}
    \mathcal{L}_{ce} = - \sum_{c=0}^{1} y_c \log(p_c),
\end{equation}

where $y_c$ denotes the ground-truth label for class $c$ (normal or anomalous) and $p_c$ represents the probability of class $c$, obtained by applying the softmax function to the logit vector $\mathbf{z}$.

\begin{equation}
\label{eq3}
p_c = \frac{\exp(z_c)}{\sum_{i=0}^{1} \exp(z_i)}.
\end{equation}

Once the training is complete, the anomaly score for a test sample $x$ is defined as the softmax probability corresponding to the anomalous class ($c=1$).

\begin{equation}
\label{eq4}
S_{LP}(x) = p_1 = \frac{\exp(z_1)}{\exp(z_0) + \exp(z_1)}.
\end{equation}

Herein, we construct two distinct data configurations for the LP evaluation to assess performance under different conditions. Specifically, the first configuration, termed the \textit{in-domain LP}, measures generalization performance on unseen data within the same domain. Each section is split equally, which yields 100 samples for classifier training and 100 for evaluation. The DCASE 2022 configuration therefore uses 300 training and 300 evaluation samples across three sections. The second configuration, termed the \textit{out-domain LP}, evaluates robustness against domain shifts. This setup adopts a leave-one-out cross-validation approach, i.e., the classifier is trained on 400 samples from two sections and evaluated on 200 samples from the remaining unseen target section. Pump and ToyConveyor use the same in-domain split and leave-one-ID-out protocol in proportion to their native per-ID test sets.

LP is used only for retrospective representation evaluation. Its classifier requires labeled normal and anomalous samples and is not a normal-only deployment rule. The MD score is fitted using normal training data alone, although its reported AUC is computed on labeled evaluation data.

\subsubsection{Mahalanobis Distance}

Unlike LP, which requires a labeled subset of the evaluation data, the MD is a non-parametric method that relies solely on the statistical distribution of the normal training data. MD, owing to its computational simplicity and effectiveness, is widely adopted in anomaly detection and out-of-distribution detection tasks.

To calculate MD, we first compute mean vector $\boldsymbol{\mu}_{\text{normal}}$ and covariance matrix $\boldsymbol{\Sigma}_{\text{normal}}$ from feature vectors $\mathbf{h}=F(x)$ of all normal training samples. During the evaluation phase, anomaly score $S_{\text{MD}}$ for a test input $x_{\text{test}}$ with feature vector $\mathbf{h}_{\text{test}}$ is calculated as the MD with respect to the learned normal distribution.

\begin{equation}
\label{eq5}
\begin{split}
    S_{\text{MD}}(x_{\text{test}}) &= (\mathbf{h}_{\text{test}} - \boldsymbol{\mu}_{\text{normal}})^\top \\
    &\quad \times \boldsymbol{\Sigma}^{-1}_{\text{normal}} (\mathbf{h}_{\text{test}} - \boldsymbol{\mu}_{\text{normal}}),
\end{split}
\end{equation}

where $(\cdot)^\top$ denotes the transpose operation and $\boldsymbol{\Sigma}^{-1}_{\text{normal}}$ represents the inverse of the covariance matrix.
Although computing domain-specific statistics can enhance detection performance, the primary objective of this study is to evaluate the intrinsic quality of the representations without relying on auxiliary information. Therefore, we refrain from using domain labels and instead compute a single set of global statistics from the entire normal training dataset for the evaluation.

\subsection{Method Used to Evaluate Proxy Tasks}
\label{sec43}

This study primarily aims to verify whether performance improvements in proxy tasks effectively translate to the acquisition of representations suitable for ASD. For this, we evaluate the extent to which each neural network has learned its designed proxy task using standard performance metrics. The scope of this evaluation encompasses AE, classification, source separation, contrastive learning, and representations from pre-trained models. Specifically for pre-trained models, as we use off-the-shelf models trained for AudioSet classification, we adopt the mAP reported in the original study as the proxy task performance metric. The proxy metric may differ from the training loss. The AE and task-specific separation networks are trained using MSE but evaluated using MAE and SI-SDR, respectively.

\subsubsection{AutoEncoder}
\label{sec431}

The objective of an AE is to compress an input signal into a latent representation and subsequently reconstruct the original signal. In the context of ASD, the AE is expected to yield considerably higher reconstruction errors for anomalous than those for normal data, as the distribution of anomalies differs from that of the normal training data.

Herein, we employ log-mel spectrograms as both input and output. The reconstruction quality is evaluated based on the MAE. The MAE represents the L1 distance between the original and reconstructed spectrograms, normalized by the total number of time-frequency bins ($N$), which can be interpreted as the average error in decibels (dB).

\begin{equation}
    \label{eq6}
    {\text{MAE}} = \frac{1}{N}\sum^{N}_{i=1} \left| M^{(i)}_{\text{in}} - M^{(i)}_{\text{out}} \right|,
\end{equation}

where $N$ denotes the total number of time-frequency bins, and $M^{(i)}_{\text{in}}$ and $M^{(i)}_{\text{out}}$ represent the $i$-th bin of the input and output mel-spectrograms, respectively.

\subsubsection{Classification}
The classification-based proxy task trained the neural network to distinguish between distinct classes, such as specific machine IDs defined for each machine type. To evaluate classification performance, we employ the Macro-$F_1$ score, calculated as the arithmetic mean of the class-wise $F_1$-scores. The $F_1$-score is defined as the harmonic mean of precision and recall:

\begin{equation}
\label{eq7}
F_1 = 2 \cdot \frac{\text{Precision} \cdot \text{Recall}}{\text{Precision} + \text{Recall}}.
\end{equation}

When the network is trained using cross-entropy loss, the final output represents class probabilities, allowing for the direct derivation of classification metrics. However, metric learning approaches that operate in the feature space, such as angular margin loss (e.g., ArcFace), do not inherently produce probability outputs suitable for immediate classification scoring. Therefore, to evaluate the classification performance of models trained using angular margin loss, we attach a single linear classifier to the pre-trained feature extractor. With the weights of the backbone network frozen, we train only this auxiliary linear layer to predict class labels and compute the $F_1$-score.

\subsubsection{Source Separation}
\label{sec433}
The objective of the source separation task is to isolate target acoustic signal $x_{\text{target}}$ from mixture signal $x_{\text{mix}}$. To evaluate separation performance, we generate evaluation mixtures by combining signals from the target class with interfering signals from non-target classes. We construct evaluation sets with signal-to-noise ratios of $-5$, $0$, and $5$~dB, generating 1,000 samples for each condition. The performance is assessed using the scale-invariant signal-to-distortion ratio (SI-SDR)~\cite{le2019sdr}. The SI-SDR is defined as follows:

\begin{equation}
\label{eq8}
\text{SI-SDR} = 10 \log_{10} \left( \frac{\| a x_{\text{target}} \|^2_2}{\| a x_{\text{target}} - \hat{x}_{\text{target}} \|^2_2} \right),
\end{equation}

where $a$ denotes the optimal scaling factor that minimizes the L2 error between the target and estimated signal and $\hat{x}_{\text{target}}$ represents the signal estimated by the neural network.

\subsubsection{Contrastive learning}

Contrastive learning aims to acquire representations by minimizing the distance between positive pairs (augmented views of the same sample) while maximizing the distance between negative pairs (views of different samples). To assess the quality of the learned feature space, we adopt the alignment and uniformity metrics proposed in~\cite{wang2020understanding}.
Alignment quantifies the closeness of feature representations for positive pairs derived from the same original sample.

\begin{equation}
\label{eq9}
\mathcal{L}_{\text{align}}(f; \alpha) = \mathbb{E}_{(x,y) \sim P_{\text{pos}}} \left[ \| f(x) - f(y) \|^\alpha_2 \right], \quad \alpha > 0,
\end{equation}

where $f(x)$ denotes the $L_2$-normalized feature vector ($f(x) = F(x) / \|F(x)\|_2$) mapped to the unit hypersphere, and $P_{\text{pos}}$ denotes the distribution of positive pairs. Uniformity measures how uniformly the feature vectors are distributed across the hypersphere.

\begin{equation}
\label{eq10}
\begin{split}
    \mathcal{L}_{\text{uniformity}}(f; t) &= \log \mathbb{E}_{x, y \overset{\text{i.i.d.}}{\sim} P_{\text{data}}} \left[ \exp \left( -t \| f(x) - f(y) \|^2_2 \right) \right], \\
    &\quad t > 0,
\end{split}
\end{equation}

where $t$ is a scaling parameter and $x, y$ are independently sampled from data distribution $P_{\text{data}}$.

\section{Experimental Results}
\label{sec5}

\subsection{Representative Proxy Task Results}
\label{sec51}
This section compares intrinsic proxy metrics with ASD performance across the nine machine types. The metrics are AE reconstruction MAE, classification macro-$F_1$, source-separation SI-SDR, contrastive alignment and uniformity, and reported AudioSet AS-2M mAP. Frozen representations are evaluated based on in-domain LP AUC, out-domain LP AUC, and MD AUC. LP measures linear anomaly separability, while MD scoring measures the distance from the normal feature distribution.

The metrics computed on the study datasets are averaged over the nine machine types unless stated otherwise. AS-2M mAP is an external benchmark score and is not averaged. The best value within each comparison group is boldfaced.

\subsubsection{AutoEncoder results}
\label{sec511}
\begin{table}[t]
\centering
\caption{Comparison of the Proxy Task and ASD performances based on AE configuration.}
\label{tab:ae_results}
\resizebox{\columnwidth}{!}{%
\begin{tabular}{ccccc}
\toprule
\textbf{Model config} & \textbf{Test MAE($\downarrow$)} & \multicolumn{2}{c}{\textbf{Linear probe (AUC \%)}} & \textbf{Mahalanobis($\uparrow$)} \\
\cmidrule(lr){3-4}
\textbf{(latent, hidden)} & dB & \textbf{In-domain($\uparrow$)} & \textbf{Out-domain($\uparrow$)} & \textbf{(AUC\%)} \\
\midrule
4, 64   & 2.70 & 76.75 & 61.78 & 58.00 \\
4, 128  & 2.61 & 75.96 & \textbf{63.85} & 58.38 \\
4, 256  & 2.56 & 76.77 & 61.77 & 58.91 \\
\midrule
8, 64   & 2.65 & 75.22 & 60.50 & 57.34 \\
8, 128  & 2.54 & 75.50 & 61.71 & 58.65 \\
8, 256  & 2.48 & 76.53 & 61.74 & 59.05 \\
\midrule
16, 64  & 2.63 & 75.70 & 59.97 & 57.45 \\
16, 128 & 2.52 & 76.31 & 61.98 & 58.26 \\
16, 256 & \textbf{2.44} & \textbf{76.91} & 61.70 & \textbf{59.22} \\
\bottomrule
\end{tabular}%
}
\end{table}

Table~\ref{tab:ae_results} summarizes the results across nine configurations. Overall, lower reconstruction errors did not
consistently translate into stronger downstream ASD performance.

The in-domain LP AUC generally increased with the hidden dimension but not monotonically. The (16, 256) model had the lowest MAE of 2.44 and the highest AUC of 76.91\%. The (4, 256) model had the second-highest AUC of 76.77\%, although four configurations had a lower MAE.

The out-domain LP AUC exhibited no monotonic relationship with the MAE. The models with the lowest and highest MAEs—(16, 256) and (4, 64), respectively—produced similar
AUCs of 61.70\% and 61.78\%, respectively. The highest AUC was 63.85\%, achieved by (4, 128). The total range was 3.88 percentage points.
MD AUC increased with the hidden dimension within each latent-dimension group. The ordering across latent dimensions was non-monotonic. Therefore, MAE alone did not explain the MD ranking.

\subsubsection{Classification Result}
\label{sec512}

\begin{table}[t]
\centering
\caption{Comparison of the Proxy Task and ASD performances based on classification network configuration and loss function.}
\label{tab:clf_results}
\resizebox{\columnwidth}{!}{%
\begin{tabular}{ccccc}
\toprule
\multicolumn{1}{c}{\textbf{Model config}} & \textbf{$F_1$-Score($\uparrow$)} & \multicolumn{2}{c}{\textbf{Linear probe (AUC\%)}} & \textbf{Mahalanobis($\uparrow$)} \\
\cmidrule(lr){3-4}

\multicolumn{1}{c}{\textbf{(backbone)}} & \textbf{(macro \%)} & \textbf{In-domain($\uparrow$)} & \textbf{Out-domain($\uparrow$)} & \textbf{(AUC\%)} \\
\midrule

\multicolumn{5}{c}{\textbf{Loss function: ArcFace}} \\
\midrule

ResNet-18  & 96.62 & \textbf{66.49} & 60.47 & 55.62 \\
ResNet-34  & 96.77 & 61.94 & 59.45 & 56.55 \\
ResNet-50  & 96.23 & 63.68 & 57.67 & 55.05 \\
ResNet-101 & \textbf{96.80} & 64.36 & 58.81 & \textbf{56.62} \\
ResNet-152 & 96.70 & 65.96 & \textbf{60.73} & 55.69 \\
\midrule

\multicolumn{5}{c}{\textbf{Loss function: Cross-Entropy}} \\
\midrule

ResNet-18  & 96.98 & \textbf{74.67} & \textbf{62.32} & 61.04 \\
ResNet-34  & 96.99 & 68.51 & 61.34 & 61.41 \\
ResNet-50  & 97.17 & 70.32 & 58.98 & 59.77 \\
ResNet-101 & 97.37 & 71.41 & 60.66 & 61.44 \\
ResNet-152 & \textbf{97.55} & 69.77 & 59.98 & \textbf{62.98} \\

\bottomrule
\end{tabular}%
}
\end{table}

Table~\ref{tab:clf_results} lists the results across ten configurations. 
Macro-$F_1$ ranged from 96.23\% to 97.55\%. This narrow range provided limited resolution for ranking downstream ASD performance.

For matched backbones, the cross-entropy exceeded ArcFace on all three ASD metrics except the out-domain LP with ResNet-152, where ArcFace and cross-entropy achieved values of 60.73\% and 59.98\%, respectively. Similar proxy scores therefore did not imply similar ASD performance.
Among the cross-entropy models, ResNet-18 exhibited the highest in-domain and out-domain LP AUCs, while ResNet-152 achieved the highest MD AUC. Thus, increasing depth did not yield consistent ASD gains.

\FloatBarrier
\subsubsection{Source Separation Results}
\label{sec513}

\begin{table}[t]
\centering
\caption{Comparison of the Proxy Task and ASD performances based on the network configuration of the Source Separation. Displayed SI-SDR values are rounded to two decimal places, whereas the correlation analyses use the corresponding full-precision values.}
\label{tab:sep_results}
\resizebox{\columnwidth}{!}{%
\begin{tabular}{ccccc}
\toprule
\multicolumn{1}{c}{\textbf{Model config}} & \textbf{SI-SDR($\uparrow$)} & \multicolumn{2}{c}{\textbf{Linear probe (AUC\%)}} & \textbf{Mahalanobis($\uparrow$)} \\
\cmidrule(lr){3-4}

\multicolumn{1}{c}{\textbf{\makecell{(Conformer block, Channel)}}} & \textbf{(dB)} & \textbf{In-domain($\uparrow$)} & \textbf{Out-domain($\uparrow$)} & \textbf{(AUC\%)} \\
\midrule

0-block, 64-ch  & 1.43 & 60.11 & 55.87 & 58.44 \\
0-block, 128-ch & 1.22 & 62.40 & 58.85 & 58.77 \\
\midrule
1-block, 64-ch  & 1.30 & 64.84 & 56.47 & 59.02 \\
1-block, 128-ch & \textbf{3.08} & 70.18 & 62.80 & 62.38 \\
\midrule
2-block, 64-ch  & 1.46 & 67.36 & 61.07 & 60.60 \\
2-block, 128-ch & 1.94 & 71.83 & 66.81 & 62.24 \\
\midrule
4-block, 64-ch  & 0.93 & 68.69 & 65.36 & 61.91 \\
4-block, 128-ch & 2.57 & \textbf{74.24} &
\textbf{69.47} & \textbf{64.54} \\

\bottomrule
\end{tabular}%
}
\end{table}

Table~\ref{tab:sep_results} lists the results across eight configurations. At every block depth, the 128-channel model outperformed the 64-channel model across all three ASD metrics. Increasing the number of Conformer blocks produced monotonic gains in the in-domain and out-domain LP AUC for each channel width. MD AUC increased monotonically for the 64-channel models and decreased only from 62.38\% to 62.24\% between the 1-block and 2-block 128-channel models.

SI-SDR was non-monotonic and reached 3.08 dB for the 1-block, 128-channel model. The 4-block, 128-channel model had a lower SI-SDR of 2.57 dB but a higher in-domain LP, a higher out-domain LP, and higher MD AUCs by 4.06, 6.67, and 2.16 percentage points, respectively. SI-SDR therefore did not preserve the ASD ranking.

The 4-block, 128-channel model outperformed the 0-block, 64-channel baseline by 14.13, 13.60, and 6.10 percentage points in terms of in-domain LP, out-domain LP, and MD AUC. It achieved the highest ASD values without the highest SI-SDR.

\subsubsection{Results Obtained from Unsupervised Contrastive Learning}
\label{sec514}
\begin{table}[t]
\centering
\caption{Comparison of the Proxy Task and ASD performances based on the network configuration of Contrastive Learning and the loss function.}

\label{tab:cont_results}
\resizebox{\columnwidth}{!}{%
\begin{tabular}{cccccc}
\toprule
\textbf{Model config} & \textbf{Alignment} & \textbf{Uniformity} & \multicolumn{2}{c}{\textbf{Linear probe (AUC\%)}} & \textbf{Mahalanobis($\uparrow$)} \\
\cmidrule(lr){4-5}

\textbf{(Backbone)} & ($\downarrow$) & ($\downarrow$) & \textbf{In-domain($\uparrow$)} & \textbf{Out-domain($\uparrow$)} & \textbf{(AUC \%)} \\
\midrule

\multicolumn{6}{c}{\textbf{Loss function: SimCLR}} \\
\midrule

ResNet-18  & 0.24 & -0.82 & \textbf{67.85} & 58.94 & \textbf{55.21} \\
ResNet-34  & 0.26 & \textbf{-0.90} & 64.05 & 53.49 & 54.77 \\
ResNet-50  & \textbf{0.13} & -0.44 & 65.74 & \textbf{60.68} & 51.96 \\
ResNet-101 & 0.17 & -0.60 & 62.80 & 55.39 & 52.32 \\
ResNet-152 & 0.21 & -0.67 & 62.06 & 52.39 & 52.30 \\
\midrule

\multicolumn{6}{c}{\textbf{Loss function: SimSiam}} \\
\midrule

ResNet-18  & 0.24 & \textbf{-0.64} & 55.95 & \textbf{56.28} & \textbf{53.53} \\
ResNet-34  & 0.15 & -0.51 & \textbf{56.50} & 51.28 & 51.85 \\
ResNet-50  & 0.07 & -0.32 & 52.24 & 51.48 & 52.90 \\
ResNet-101 & 0.06 & -0.16 & 49.88 & 50.69 & 46.51 \\
ResNet-152 & \textbf{0.03} & -0.14 & 49.36 & 50.23 & 48.39 \\

\bottomrule
\end{tabular}%
}
\end{table}

Table \ref{tab:cont_results} lists the results across ten configurations.

Both methods transferred weakly and inconsistently. MD AUC ranged from 46.51\% to 55.21\%, while the maximum LP AUCs were 67.85\% in-domain and 60.68\% out-domain.
For SimSiam, the alignment decreased from 0.24 to 0.03, whereas the uniformity increased from -0.64 to -0.14 between ResNet-18 and ResNet-152. Better alignment corresponded to poorer uniformity and lower ASD performance. This
pattern is consistent with increasing representation concentration.
SimCLR displayed no monotonic trend. ResNet-34 had the lowest uniformity value of -0.90, while ResNet-50 had the lowest alignment value of 0.13 but a higher uniformity value of -0.44. Neither diagnostic consistently tracked ASD performance.
Collapse-like behavior was evident for larger task-specific SimSiam backbones but not as a common pattern across both methods. Under the evaluated
 data and augmentations, alignment and uniformity did not reliably identify representations that transferred well to ASD.

\FloatBarrier
\subsubsection{Pre-trained Model Results}
\label{sec:pretrained_results}

\begin{table}[t]
\centering
\caption{Comparison of the Proxy Task and ASD performances based on the different pre-trained model configurations.}
\label{tab:pre_results}
\resizebox{\columnwidth}{!}{%
\begin{tabular}{ccccc}
\toprule
\multicolumn{1}{c}{\textbf{Model config}} & \textbf{AS-2M($\uparrow$)} & \multicolumn{2}{c}{\textbf{Linear probe (AUC \%)}} & \textbf{Mahalanobis($\uparrow$)} \\
\cmidrule(lr){3-4}

\multicolumn{1}{c}{} & \textbf{mAP (\%)} & \textbf{In-domain($\uparrow$)} & \textbf{Out-domain($\uparrow$)} & \textbf{(AUC\%)} \\
\midrule

EAT-large    & 49.5 & 74.00 & 56.97 & 63.44 \\
EAT-base     & 48.9 & 73.93 & 58.87 & 64.11 \\
\midrule
BEATs-iter3  & 48.0 & 75.80 & 59.74 & 65.23 \\
BEATs-iter3+ & 48.6 & \textbf{77.64} & \textbf{61.40} & \textbf{66.37} \\
\midrule
CED-base & \textbf{50.0} & 58.85 & 54.86 & 57.67 \\
CED-mini & 49.0 & 57.42 & 53.35 & 57.14 \\
CED-small & 49.6 & 58.10 & 54.49 & 57.92 \\
CED-tiny & 48.1 & 59.99 & 53.85 & 57.27 \\

\bottomrule
\end{tabular}%
}
\end{table}

Table~\ref{tab:pre_results} lists the results across eight model variants. AS-2M mAP did not rank ASD performance consistently. BEATs-iter3+ had the highest three ASD AUCs despite attaining the third-lowest mAP. EAT-base had a lower mAP than EAT-large but a higher out-domain LP and MD AUC. Their in-domain LP values were similar.
CED-base had the highest mAP of 50.0\%, but every CED variant had lower AUCs than the EAT and BEATs variants. The mAP of BEATs-iter3+ was 1.4 percentage points lower than that of CED-base, while its in-domain LP, out-domain LP, and MD AUCs were higher by 18.79, 6.54, and 8.70 percentage points.
Under the frozen-encoder protocol, AudioSet tagging mAP was not a reliable selection criterion for ASD representations.

\subsection{Shared-Backbone Results}
\label{sec53}

\begin{table}[!htbp]
\centering
\caption{Shared-backbone results averaged over the nine machine types.
All trainable proxy-task models use the same Lite encoder family
instantiated at five capacity levels (Lite~0--Lite~4), while retaining
task-specific adapters and prediction heads. The best value(s) within
each task block are shown in bold.}
\label{tab:shared_results}
\resizebox{\columnwidth}{!}{%
\begin{tabular}{ccccc}
\toprule
\multicolumn{1}{c}{\textbf{Model config}} & \textbf{Task metric / diagnostic} & \multicolumn{2}{c}{\textbf{Linear probe (AUC\%)}} & \textbf{Mahalanobis distance AUC} \\
\cmidrule(lr){3-4}
\multicolumn{1}{c}{\textbf{(backbone)}} & \textbf{(task-specific)} & \textbf{In-domain($\uparrow$)} & \textbf{Out-domain($\uparrow$)} & \textbf{(AUC\%)} \\
\midrule

\multicolumn{5}{c}{\textbf{AutoEncoder --- proxy: reconstruction $L_1$ ($\downarrow$)}} \\
\midrule
Lite0 & \textbf{3.29} & 65.11 & 54.78 & 59.12 \\
Lite1 & 3.60 & 67.79 & 55.38 & 58.93 \\
Lite2 & 3.63 & 65.82 & 55.08 & 58.50 \\
Lite3 & 3.76 & 66.83 & 58.19 & \textbf{59.45} \\
Lite4 & 5.46 & \textbf{69.38} & \textbf{59.14} & 58.91 \\
\midrule

\multicolumn{5}{c}{\textbf{Classification: ArcFace --- proxy: macro-$F_1$ (\%, $\uparrow$)}} \\
\midrule
Lite0 & \textbf{96.66} & 72.58 & \textbf{64.27} & 60.19 \\
Lite1 & 96.43 & 72.07 & 61.54 & 62.61 \\
Lite2 & \textbf{96.66} & \textbf{74.00} & 63.87 & 63.25 \\
Lite3 & 96.30 & 70.75 & 63.62 & 63.26 \\
Lite4 & 96.20 & 72.94 & 60.97 & \textbf{64.85} \\
\midrule

\multicolumn{5}{c}{\textbf{Classification: Cross-Entropy --- proxy: macro-$F_1$ (\%, $\uparrow$)}} \\
\midrule
Lite0 & 96.53 & \textbf{75.15} & 65.95 & 64.04 \\
Lite1 & 95.62 & 72.85 & 65.00 & 63.63 \\
Lite2 & \textbf{96.56} & 73.55 & 65.38 & 65.37 \\
Lite3 & 94.54 & 74.18 & \textbf{67.36} & 64.14 \\
Lite4 & 96.29 & 73.89 & 66.69 & \textbf{67.16} \\
\midrule

\multicolumn{5}{c}{\textbf{Source separation (Mel-domain adapter)
--- proxy: component-reconstruction $L_1$ ($\downarrow$)}} \\
\midrule
Lite0 & \textbf{4.16} & 61.45 & 52.87 & \textbf{59.77} \\
Lite1 & 4.59 & 63.17 & \textbf{57.45} & 59.66 \\
Lite2 & 4.28 & 62.98 & 54.72 & 59.55 \\
Lite3 & 4.41 & 63.18 & 54.66 & 59.45 \\
Lite4 & 4.75 & \textbf{64.53} & 56.26 & 59.34 \\
\midrule

\multicolumn{5}{c}{\textbf{Contrastive: SimCLR --- diagnostics:
uniformity / alignment ($\downarrow$)}} \\
\midrule
Lite0 & \textbf{-2.08} / 0.56 & \textbf{64.72} & \textbf{57.18} & 57.12 \\
Lite1 & -2.05 / 0.53 & 62.98 & 56.34 & 56.14 \\
Lite2 & -2.02 / 0.51 & 63.42 & 55.78 & 57.11 \\
Lite3 & -1.97 / 0.48 & 63.02 & 56.87 & 57.43 \\
Lite4 & -1.97 / \textbf{0.46} & 61.77 & 56.96 & \textbf{57.62} \\
\midrule

\multicolumn{5}{c}{\textbf{Contrastive: SimSiam --- diagnostics:
uniformity / alignment ($\downarrow$)}} \\
\midrule
Lite0 & -0.91 / 1.66 & 56.95 & \textbf{53.42} & 49.59 \\
Lite1 & -1.14 / \textbf{1.36} & 55.70 & 51.41 & 49.47 \\
Lite2 & \textbf{-1.21} / 1.39 & 54.69 & 52.78 & 51.80 \\
Lite3 & -0.85 / 1.61 & \textbf{57.21} & 51.49 & 49.04 \\
Lite4 & -0.92 / 1.62 & 53.88 & 50.18 & \textbf{51.98} \\
\bottomrule
\end{tabular}%
}
\end{table}

The shared-backbone experiment addresses two questions: (i) whether intrinsic proxy metrics rank model variants more consistently with ASD performance within a common encoder family and (ii) whether relative transfer patterns across proxy-task families persist after reducing between-task architectural variation. Each trainable proxy task is instantiated with EfficientNet-Lite0--Lite4 encoders while retaining task-specific heads and adapters. The analysis therefore standardizes the encoder family rather than the complete task pipeline. The pre-trained model family is excluded because its encoders are not trained under this configuration. Section~\ref{sec:62} compares these results with those from the task-specific experiments.

Classification yielded the highest downstream AUCs under the shared encoder family. Cross-entropy classification achieved the highest AUC among all models in this analysis in terms of in-domain LP (75.15\%, Lite0), out-domain LP (67.36\%, Lite3), and MD scoring (67.16\%, Lite4). Because the maxima occurred at different capacities, no single capacity was optimal across all three protocols. For each protocol, the maximum ArcFace AUC also exceeded the corresponding maxima for AE, separation, SimCLR, and SimSiam. SimSiam produced the lowest AUCs in all three protocols.

The intrinsic metrics did not consistently identify the configuration with the highest ASD performance within the AE, classification, or separation branches. For the AE, Lite0 had the lowest reconstruction MAE (3.29), whereas Lite4 produced the highest in-domain and out-domain LP AUCs (69.38\% and 59.14\%, respectively), while Lite3 produced the highest MD AUC (59.45\%). For cross-entropy classification, Lite2 had the highest macro-$F_1$ (96.56\%), whereas the three protocol-specific maxima were achieved by Lite0, Lite3, and Lite4. For ArcFace, Lite0 and Lite2 shared the highest displayed macro-$F_1$ (96.66\%), whereas the in-domain LP, out-domain LP, and MD maxima occurred for Lite2, Lite0, and Lite4, respectively. For the shared separation branch, Lite0 exhibited the lowest component-reconstruction MAE (4.16) and the highest MD AUC (59.77\%), whereas Lite4 and Lite1 achieved the highest in-domain and out-domain LP AUCs (64.53\% and 57.45\%), respectively. Agreement with one ASD protocol therefore did not imply agreement with the others.

For SimCLR, the lowest uniformity and the lowest alignment values occurred at different capacities (Lite0 and Lite4, respectively), each coinciding with a different protocol-specific ASD maximum; therefore, the diagnostics did not order ASD performance consistently. SimSiam showed no monotonic proxy–ASD pattern. The shared separation branch is not directly equivalent to the task-specific STFT separator, because it uses mel-domain component reconstruction and a different proxy metric. Section 6.2 therefore compares the two settings in terms of ranking behavior and transfer patterns rather than numerical proxy values. Standardizing the encoder family did not yield stable proxy-based rankings, whereas cross-entropy continued to produce comparatively strong ASD representations.

\subsection{Checkpoint Sensitivity Analysis}
\label{sec54}

\begin{figure}
    \centering
    \includegraphics[width=1.0\linewidth]{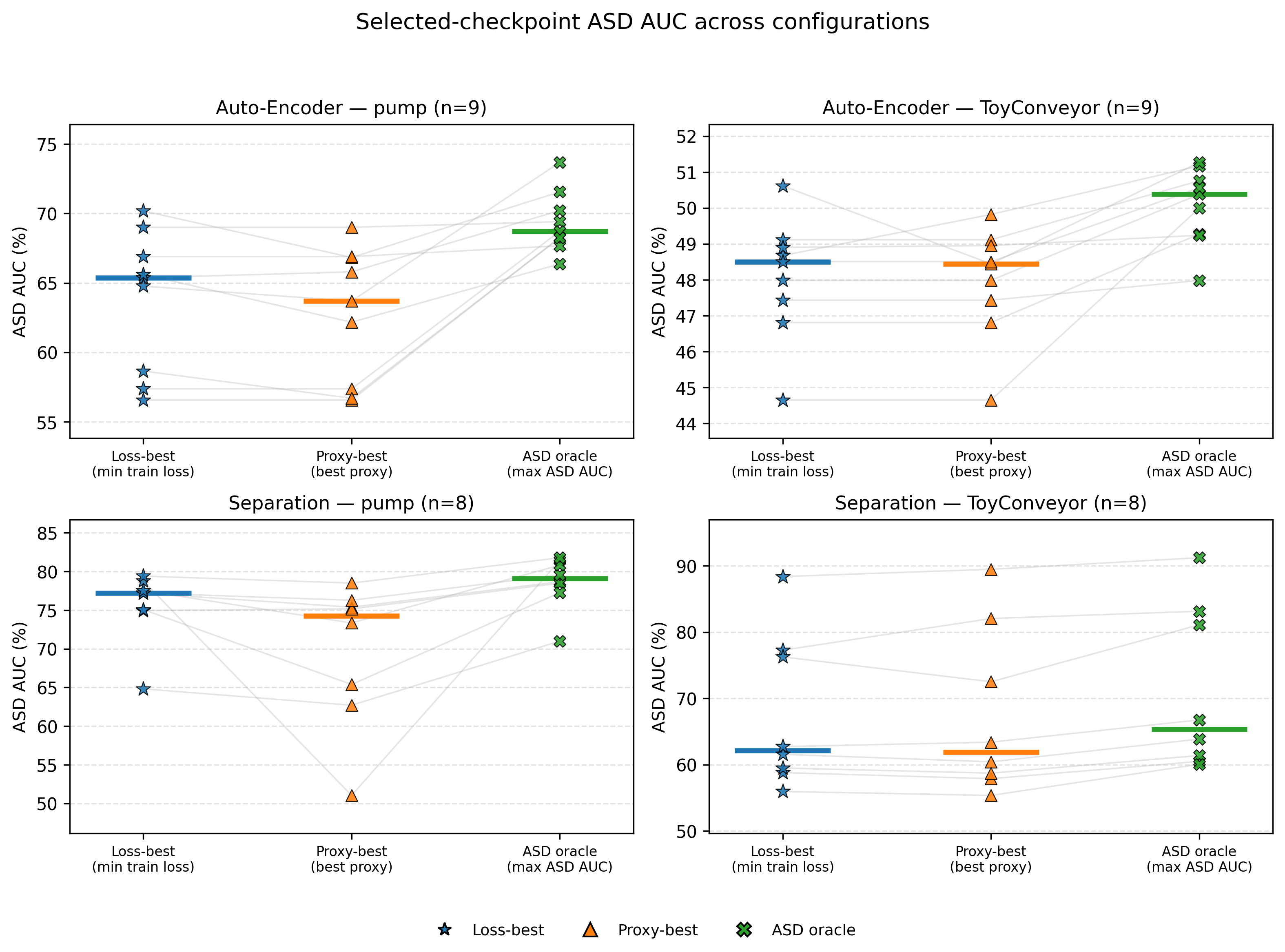}
    \caption{Out-domain LP AUC under loss-best, proxy-best, and out-domain-oracle checkpoint selection for the AE and source-separation models on Pump and ToyConveyor. The markers denote model configurations, the gray lines connect matched configurations, and the horizontal bars indicate medians. The vertical-axis limits differ across panels.}
    \label{fig:checkpoint}
\end{figure}

For the main experiments, we select the minimum-training-loss checkpoint and use no anomaly-labeled validation data. We compare this rule with two post-hoc references: loss-best minimizes the training loss, while proxy-best minimizes the test MAE for AE or maximizes the mean test SI-SDR across -5, 0, and 5 dB for source separation. The out-domain oracle maximizes the mean section-wise out-domain LP AUC among the saved checkpoints. The analysis includes nine AE and eight source-separation configurations for Pump and ToyConveyor.

Loss-best had a higher median out-domain LP AUC than proxy-best in three of the four panels. The medians were 65.37\% versus 63.70\% for AE–Pump, 48.51\% versus 48.45\% for AE–ToyConveyor, and 77.20\% versus 70.06\% for separation–Pump. For separation–ToyConveyor, the proxy-best and loss-best were 62.40\% and 62.13\%, respectively. Proxy-best therefore showed no systematic advantage. The median configuration-wise gap to the oracle ranged from 2.06 to 4.84 percentage points for loss-best and from 2.06 to 6.59 points for proxy-best. The largest gap was 30.39 points for proxy-best in separation–Pump. The proxy–ASD mismatch was therefore not explained solely by the checkpoint selection under these settings.

\FloatBarrier
\section{Discussion}
\label{sec6}
\subsection{Association Between Proxy Quality and ASD Performance}

\begin{figure}[!htbp]
\centering
\includegraphics[width=0.60\textwidth]{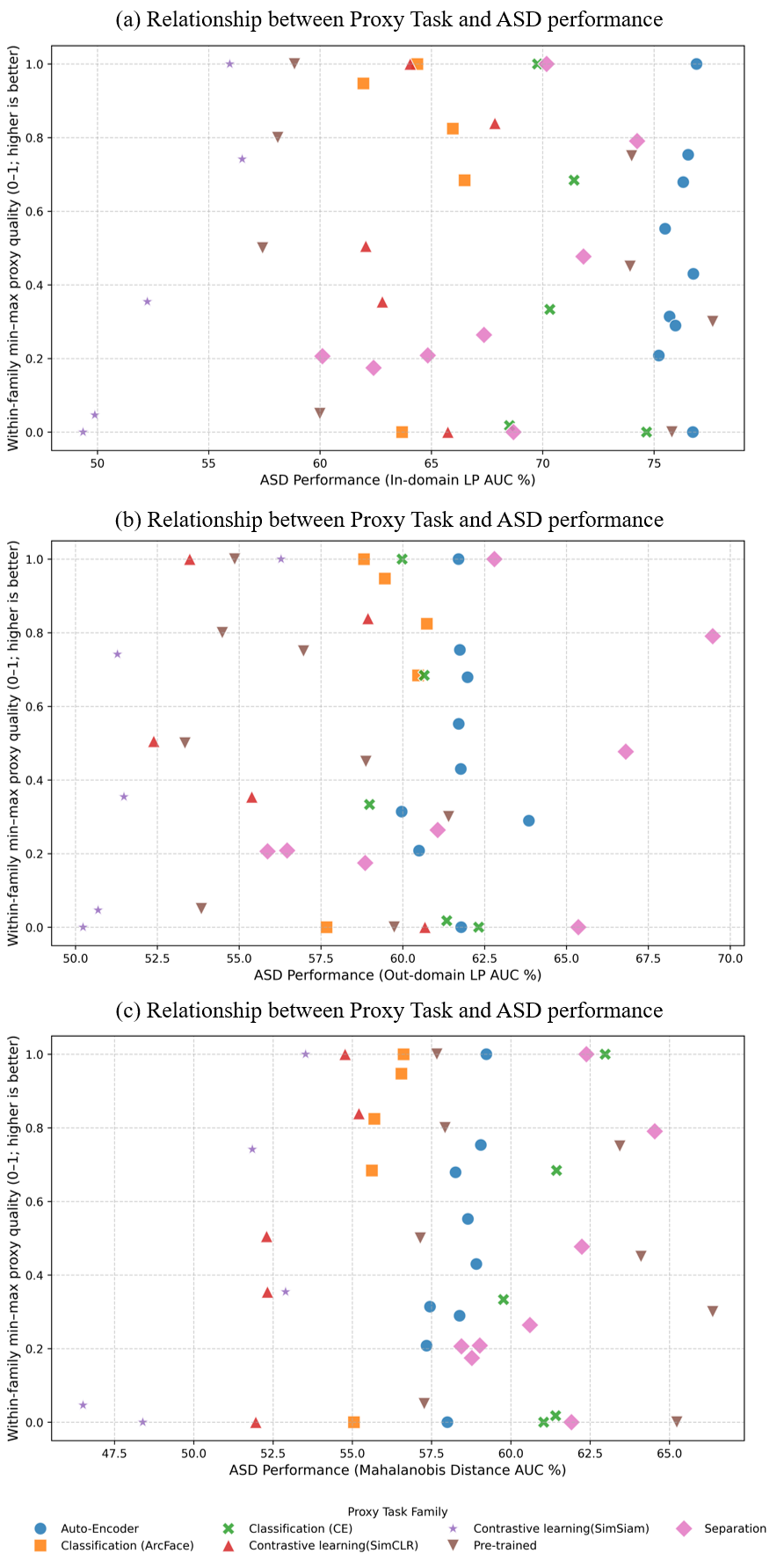}
\caption{Within-family associations between direction-adjusted proxy quality and ASD performance for the task-specific implementations: (a) in-domain LP, (b) out-domain LP, and (c) MD-based AUC. Proxy values are min--max normalized separately within each family and are therefore not directly comparable across families. Marker shape and color denote the proxy family; Spearman correlations are reported in Table~\ref{tab:correlation_summary}.}
\label{fig:corr}
\end{figure}

Table~\ref{tab:correlation_summary} lists the within-family Spearman correlations for the task-specific models, while Figure~\ref{fig:corr} shows the corresponding scatter plots. No association remained significant after a Benjamini--Hochberg correction across the 21 exploratory tests. Because each family contains only five to nine designed configurations, the correlations and permutation $p$-values are descriptive. Most families also form capacity ladders. A common change in capacity can therefore move the proxy and ASD metrics in the same direction without establishing proxy--target alignment.

The main evidence concerns rank preservation. The highest SI-SDR configuration was not the best under any ASD protocol. The lowest AE reconstruction error did not correspond to the highest out-domain LP AUC. The highest classification score and the highest AudioSet mAP did not identify the best representation across all three protocols. These reversals show that the evaluated proxy metrics did not provide a stable model ranking.

All positive coefficients are interpreted under the same criterion. They may reflect common ordering by capacity or feature dimension. This issue is relevant to the ResNet families because the representation dimension changes from 512 for ResNet-18 and ResNet-34 to 2048 for deeper models. Capacity and feature dimension are therefore confounded. The latter can also affect covariance estimation for MD scoring. The high coefficients observed for ArcFace and SimSiam do not provide stronger evidence of proxy alignment than the lower coefficients observed for AE or source separation.

The pre-trained coefficients are not interpreted as evidence of a general negative relation between AudioSet mAP and ASD transfer. The observed mAP range is 48.0–50.0\%, and the encoders differ in their objective, architecture, pooling, and feature dimension.

Proxy-metric ranking and objective transfer remain distinct. Macro-$F_1$ did not provide a stable ranking, whereas cross-entropy representations transferred comparatively well in the task-specific and shared-encoder analyses.

\begin{table}[!htbp]
\centering
\caption{Within-family Spearman rank correlations ($\rho$) between direction-adjusted proxy quality and ASD performance for the task-specific configurations. MAE and uniformity were sign-reversed so that larger values indicate higher proxy quality. Exact paired-permutation tests were treated as exploratory. Two associations were nominally significant before correction, and neither remained significant after Benjamini--Hochberg correction across the 21 tests (minimum $q=0.181$).}
\label{tab:correlation_summary}
\resizebox{\columnwidth}{!}{%
\begin{tabular}{lccc}
\toprule
\textbf{Proxy task} & \textbf{In-domain LP} & \textbf{Out-domain LP} & \textbf{MD} \\
\midrule
AutoEncoder & $0.40$ & $-0.10$ & $0.78$ \\
Classification (CE) & $-0.30$ & $-0.70$ & $0.70$ \\
Classification (ArcFace) & $-0.20$ & $0.10$ & $1.00$ \\
Source separation & $0.62$ & $0.43$ & $0.67$ \\
Contrastive (SimCLR) & $0.10$ & $-0.50$ & $0.80$ \\
Contrastive (SimSiam) & $0.90$ & $0.90$ & $0.80$ \\
Pre-trained & $-0.57$ & $-0.36$ & $-0.38$ \\
\bottomrule
\end{tabular}%
}
\end{table}

\FloatBarrier
\subsection{Proxy-Metric Ranking, Objective Transfer, and Task-Interface Dependence}
\label{sec:62}

Here, the task interface denotes the task-specific head, adapter, output formulation, and representation provided to the downstream ASD scorer.

A comparison between the results from the task-specific and shared-backbone experiments highlights two findings. First, between-task architectural variation is not sufficient for explaining the proxy–ASD mismatch. In the task-specific experiments, the configuration with the best proxy metric frequently differed from the configuration with the best ASD performance, and the same pattern persisted when the trainable proxy tasks were restricted to the EfficientNet-Lite family. The AE with the lowest MAE and the cross-entropy classifier with the highest macro-$F_1$ did not maximize any of the three ASD metrics, and the relationships of the separation and contrastive diagnostics with ASD performance remained protocol-dependent. This experiment did not isolate the proxy objective, because the encoder capacity still varied within the family, task-specific heads, and adapters, while the output formulations differed across objectives, and the AE and separation branches used different downstream representations under the two settings. Nevertheless, the mismatch persisted under a common encoder family, which indicates that it is not solely an artifact of differing encoder architectures.

Second, proxy-metric ranking and objective transfer are empirically distinct. Cross-entropy classification produced comparatively strong ASD representations in both the task-specific ResNet and shared EfficientNet-Lite experiments, although macro-$F_1$ did not rank the model variants consistently under either setting. An objective can therefore yield useful ASD representations even when its intrinsic metric is a weak criterion for model selection.

The source-separation transfer differed substantially between the two implementations. The task-specific STFT-domain Conformer implementation achieved maximum in-domain LP, out-domain LP, and MD AUCs of 74.24\%, 69.47\%, and 64.54\%, respectively. The corresponding maxima for the shared mel-domain branch were 64.53\%, 57.45\%, and 59.77\%. Under the task-specific setting, SI-SDR exhibited positive descriptive correlations with all three ASD metrics, but the configuration with the highest SI-SDR was not the best under any ASD protocol. Under the shared setting, the configuration with the lowest MAE also had the highest MD AUC but not the highest LP AUCs, while MD AUC varied by only 0.43 percentage points across capacities. These results do not support a consistent association between separation quality and ASD transfer across implementations. Because the settings also differ in representation domain, decoder, decomposition target, downstream feature interface, and proxy metric, the lower AUCs of the shared branch cannot be attributed to the encoder change alone.

The contrastive results also varied across model designs. The capacity-dependent concentration pattern observed for task-specific SimSiam was not reproduced within the shared encoder family. This diagnostic pattern was therefore not stable across implementations. SimCLR was also non-monotonic, providing no consistent capacity-based explanation. These experiments motivate separate evaluations for three questions: whether an intrinsic proxy metric ranks model variants consistently with ASD performance, whether the proxy objective yields representations that transfer to ASD, and whether the ranking and transfer patterns remain robust to changes in the encoder and task interface.

\subsection{Evaluation-Dependent Representation Properties}
\label{projection_section}

The representation rankings varied by evaluation protocol. LP measures the linear separation between normal and anomalous samples, while MD scoring measures the distance from the normal feature distribution. The protocols therefore selected different configurations.

AE reconstruction quality was more closely associated with MD AUC than with LP. Among the cross-entropy classifiers, ResNet-18 produced the best LP values while ResNet-152 produced the best MD value. The preferred capacity also differed across the shared-backbone metrics. No single protocol reproduced all rankings.

\begin{figure}[!htbp]
    \centering
    \includegraphics[width=\linewidth]{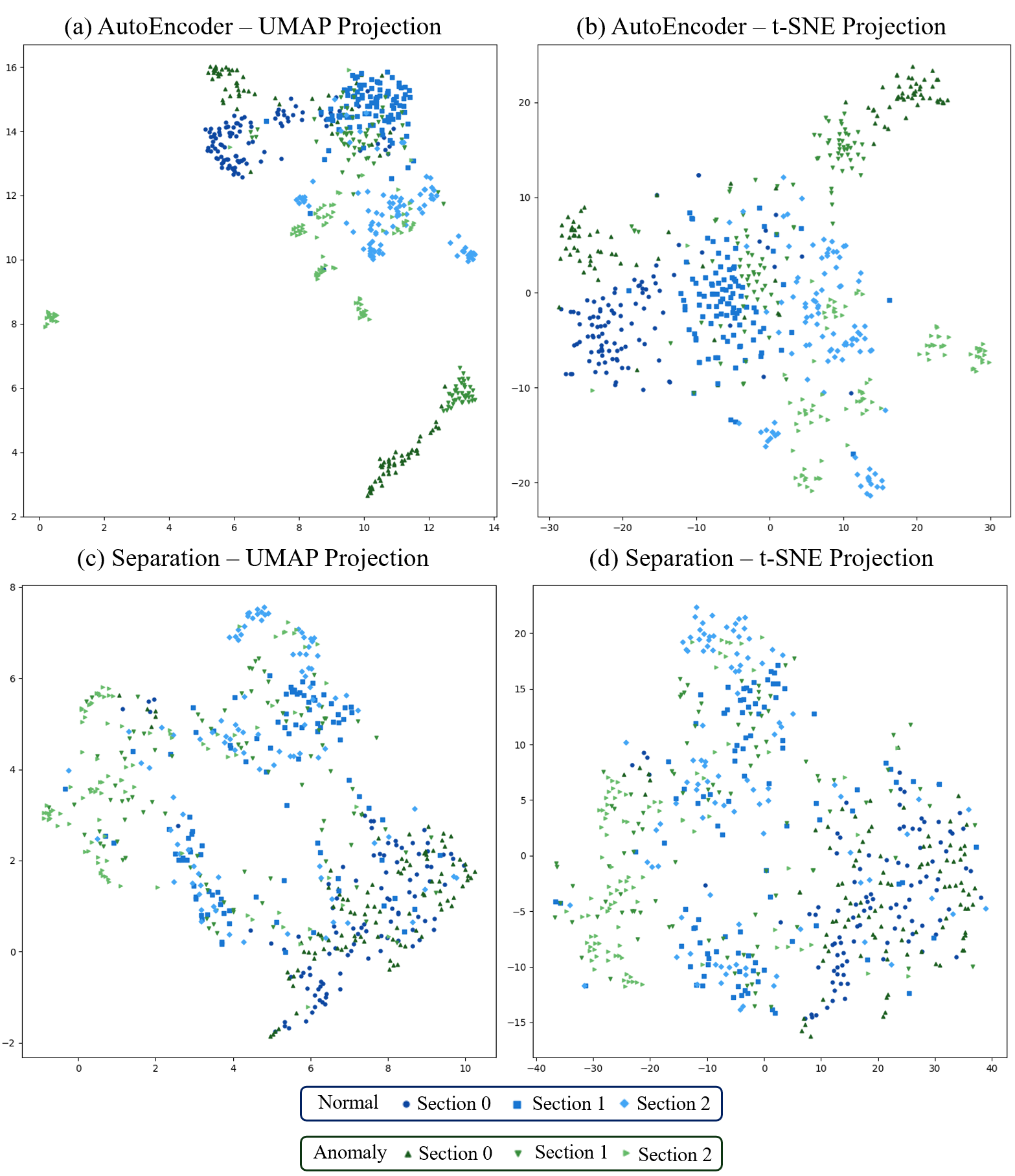}
    \caption{UMAP and t-SNE projections of the \textit{gearbox} representations: (a) AE with UMAP, (b) AE with t-SNE, (c) source separation with UMAP, and (d) source separation with t-SNE. Although the AE projections show greater apparent separation, the source-separation model achieved the higher MD-based AUC (72.68\% vs.\ 64.15\%).}
    \label{fig:5}
\end{figure}

Two-dimensional projections are qualitative diagnostics. As seen in the gearbox example in Fig.~\ref{fig:5}, the AE representation showed clearer separation in UMAP and t-SNE, but its MD AUC was 64.15\% compared with 72.68\% for source separation. Projected separation therefore did not preserve the original-space ranking.

\FloatBarrier
\subsection{Limitations}
This study covers nine machine types from two dataset families. Each within-family analysis contains five to nine designed configurations and does not provide independent random samples. The reported correlations may reflect common ordering by model capacity. In the ResNet families, capacity and feature dimension are confounded because the output dimension changes from 512 to 2048.

The pre-trained analysis covers a narrow AS-2M mAP range of 48.0–50.0\% and combines heterogeneous objectives, frontends, pooling rules, and feature dimensions. Its correlations do not establish a general relationship between AudioSet mAP and ASD transfer. The MD score also uses a single global normal distribution without section labels, which can merge section-specific normal modes.

The shared-encoder analysis controls only the encoder family. The separation branch also changes the representation domain, decoder, task interface, and proxy metric. The checkpoint analysis covers only two tasks, two machine types, and the out-domain LP. Figure~\ref{fig:5} presents an illustrative single-machine example and does not provide a systematic projection comparison.

Finally, all trainable proxy objectives were evaluated in a single-task setting. Multi-task combinations may exploit complementary representation properties across proxy objectives, but a controlled multi-task study would require separate analysis of loss balancing, task scheduling, shared/private heads, and hyperparameter search.

\subsection{Practical Implications}

Proxy-metric ranking, objective transfer, and compatibility with the encoder and task interface should be evaluated separately. A high proxy score does not guarantee an ASD-oriented representation. A weak proxy–ASD correlation also does not indicate that the objective is ineffective. Classification illustrates this distinction. Macro-$F_1$ had limited protocol-dependent ranking value, while cross-entropy representations transferred comparatively well.

The checkpoint analysis provides a limited robustness check. For AE and source separation on Pump and ToyConveyor, proxy-best selection did not systematically improve the out-domain LP AUC over minimum-loss selection. This result is limited to two tasks, two machine types, and one downstream protocol.

The resulting evaluation procedure has three stages:
\begin{itemize}
\item \textbf{Stage 1: Proxy-task health and metric resolution.} Confirm that the intrinsic metric meaningfully differentiates model variants and that training is not dominated by saturation, collapse-like behavior, or optimization failure.
\item \textbf{Stage 2: Objective transfer to ASD.} Evaluate whether the learned frozen representations support ASD under complementary downstream protocols (in-domain and out-domain linear probing and a distance-based anomaly score), independently of whether the proxy metric ranks them correctly.
\item \textbf{Stage 3: Ranking and interface robustness.} Test whether proxy--ASD rankings and objective-level conclusions persist across encoder families, capacities, task interfaces, and checkpoint-selection rules.
\end{itemize}
Proxy tasks should not be selected based on convergence or benchmark score alone. System design should evaluate the objective, architecture, task interface, and downstream protocol separately.

\section{Conclusion}
\label{sec7}

We determined whether intrinsic proxy metrics rank anomalous sound detection representations. Five families were compared across nine machine types in terms of in-domain LP, out-domain LP, and MD scoring. The evaluation also included shared-backbone and checkpoint-sensitivity analyses.

No within-family association remained significant after multiple-comparison correction. AE reconstruction quality with MD AUC ($\rho=0.78$) and ArcFace macro-$F_1$ with MD AUC ($\rho=1.00$) were nominally significant before correction, with $q=0.181$ after correction. Source separation showed positive correlations of $\rho=0.43$--$0.67$, but the highest SI-SDR did not yield the best ASD result. Macro-$F_1$ and AudioSet mAP did not rank ASD representations consistently. Ranking ability and objective usefulness were nonetheless distinct: cross-entropy classification transferred comparatively well, whereas contrastive learning transferred weakly.

With the shared EfficientNet-Lite encoders, cross-entropy classification produced the highest in-domain, out-domain, and MD AUCs of 75.15\%, 67.36\%, and 67.16\%, respectively, at different capacities. The task-specific separation result was not retained with the mel-domain adapter, but the comparison also changed the task formulation and proxy metric. Proxy-best checkpoint selection did not systematically improve out-domain LP over minimum-loss selection.

Proxy-metric ranking, objective transfer, and task-interface dependence should be evaluated separately. The proposed three-stage procedure operationalizes these distinctions, covering proxy-task health and metric resolution, objective transfer to ASD, and ranking and interface robustness.

The main limitations are the use of two dataset families, single-task training, a narrow range of source-task scores, and a shared separation formulation that differs from the task-specific STFT model. Future work should use broader industrial datasets, multi-task objectives, task-preserving architectural controls, and fine-tuned pre-trained models.
\clearpage


\bibliographystyle{elsarticle-num}
\bibliography{references}

@inproceedings{valenzise2007scream,
  author={Valenzise, G. and Gerosa, L. and Tagliasacchi, M. and Antonacci, F. and Sarti, A.},
  booktitle={2007 IEEE Conference on Advanced Video and Signal Based Surveillance}, 
  title={Scream and gunshot detection and localization for audio-surveillance systems}, 
  year={2007},
  volume={},
  number={},
  pages={21-26},
  doi={10.1109/AVSS.2007.4425280}
}

@inproceedings{clavel2005events,
  author={Clavel, C. and Ehrette, T. and Richard, G.},
  booktitle={2005 IEEE International Conference on Multimedia and Expo}, 
  title={Events Detection for an Audio-Based Surveillance System}, 
  year={2005},
  volume={},
  number={},
  pages={1306-1309},
  doi={10.1109/ICME.2005.1521669}
}

@inproceedings{clavel2008detection,
  author={Clavel, C. and Devillers, L. and Richard, G. and Vasilescu, I. and Ehrette, T.},
  booktitle={2007 IEEE International Conference on Acoustics, Speech and Signal Processing - ICASSP '07}, 
  title={Detection and Analysis of Abnormal Situations Through Fear-Type Acoustic Manifestations}, 
  year={2007},
  volume={4},
  number={},
  pages={IV-21-IV-24},
  doi={10.1109/ICASSP.2007.367153}
}

@article{foggia2015audio,
  author={Foggia, Pasquale and Petkov, Nicolai and Saggese, Alessia and Strisciuglio, Nicola and Vento, Mario},
  journal={IEEE Transactions on Intelligent Transportation Systems}, 
  title={Audio Surveillance of Roads: A System for Detecting Anomalous Sounds}, 
  year={2016},
  volume={17},
  number={1},
  pages={279-288},
  doi={10.1109/TITS.2015.2470216}
}

@inproceedings{marchi2015nonlinear,
  author={Marchi, Erik and Vesperini, Fabio and Weninger, Felix and Eyben, Florian and Squartini, Stefano and Schuller, Björn},
  booktitle={2015 International Joint Conference on Neural Networks (IJCNN)}, 
  title={Non-linear prediction with LSTM recurrent neural networks for acoustic novelty detection}, 
  year={2015},
  volume={},
  number={},
  pages={1-7},
  doi={10.1109/IJCNN.2015.7280757}
}

@inproceedings{marchi2015novel,
  author={Marchi, Erik and Vesperini, Fabio and Eyben, Florian and Squartini, Stefano and Schuller, Björn},
  booktitle={2015 IEEE International Conference on Acoustics, Speech and Signal Processing (ICASSP)}, 
  title={A novel approach for automatic acoustic novelty detection using a denoising autoencoder with bidirectional LSTM neural networks}, 
  year={2015},
  volume={},
  number={},
  pages={1996-2000},
  doi={10.1109/ICASSP.2015.7178320}
}

@inproceedings{kawachi2018complementary,
  author={Kawachi, Yuta and Koizumi, Yuma and Harada, Noboru},
  booktitle={2018 IEEE International Conference on Acoustics, Speech and Signal Processing (ICASSP)}, 
  title={Complementary Set Variational Autoencoder for Supervised Anomaly Detection}, 
  year={2018},
  volume={},
  number={},
  pages={2366-2370},
  doi={10.1109/ICASSP.2018.8462181}}

@inproceedings{koizumi2019toyadmos,
title={{ToyADMOS}: A Dataset of Miniature-Machine Operating Sounds for Anomalous Sound Detection},
  author={Koizumi, Yuma and Saito, Shoichiro and Uematsu, Hisashi and Harada, Noboru and Imoto, Keisuke},
  booktitle={2019 IEEE Workshop on Applications of Signal Processing to Audio and Acoustics (WASPAA)},
  pages={313--317},
  year={2019},
  organization={IEEE},
  doi={10.1109/WASPAA.2019.8937164}
}

@inproceedings{purohit2019mimii,
title={{MIMII} Dataset: Sound Dataset for Malfunctioning Industrial Machine Investigation and Inspection},
  author={Purohit, Harsh and Tanabe, Ryo and Ichige, Kenji and Endo, Takashi and Nikaido, Yuki and Suefusa, Kaori and Kawaguchi, Yohei},
  booktitle={Proc. Detection and Classification of Acoustic Scenes and Events 2019 Workshop (DCASE2019)},
  pages={161--165},
  year={2019},
  month={Oct.},
  address={New York, NY, USA}
}

@inproceedings{koizumi2020description,
    Author = "Koizumi, Yuma and Kawaguchi, Yohei and Imoto, Keisuke and Nakamura, Toshiki and Nikaido, Yuki and Tanabe, Ryo and Purohit, Harsh and Suefusa, Kaori and Endo, Takashi and Yasuda, Masahiro and Harada, Noboru",
    title = "Description and Discussion on {DCASE}2020 Challenge Task2: Unsupervised Anomalous Sound Detection for Machine Condition Monitoring",
    year = "2020",
    booktitle = "Proceedings of the Detection and Classification of Acoustic Scenes and Events 2020 Workshop (DCASE2020)",
    month = "November",
    pages = "81--85",
}

@inproceedings{Giri2020a,
    author = "Giri, Ritwik and Tenneti, Srikanth V. and Cheng, Fangzhou and Helwani, Karim and Isik, Umut and Krishnaswamy, Arvindh",
    title = "Self-Supervised Classification for Detecting Anomalous Sounds",
    booktitle = "Proceedings of the Detection and Classification of Acoustic Scenes and Events 2020 Workshop (DCASE2020)",
    address = "Tokyo, Japan",
    month = "November",
    year = "2020",
    pages = "46--50",
}

@inproceedings{harada2021toyadmos2,
    author = "Harada, Noboru and Niizumi, Daisuke and Takeuchi, Daiki and Ohishi, Yasunori and Yasuda, Masahiro and Saito, Shoichiro",
    title = "ToyADMOS2: Another Dataset of Miniature-Machine Operating Sounds for Anomalous Sound Detection under Domain Shift Conditions",
    booktitle = "Proceedings of the 6th Detection and Classification of Acoustic Scenes and Events 2021 Workshop (DCASE2021)",
    address = "Barcelona, Spain",
    month = "November",
    year = "2021",
    pages = "1--5",
    isbn = "978-84-09-36072-7",
    doi. = "10.5281/zenodo.5770113"
}

@article{koizumi2018unsupervised,
  title={Unsupervised detection of anomalous sound based on deep learning and the neyman--pearson lemma},
  author={Koizumi, Yuma and Saito, Shoichiro and Uematsu, Hisashi and Kawachi, Yuta and Harada, Noboru},
  journal={IEEE/ACM Transactions on Audio, Speech, and Language Processing},
  volume={27},
  number={1},
  pages={212--224},
  year={2018},
  publisher={IEEE}
}

@article{jardine2006review,
  title={A review on machinery diagnostics and prognostics implementing condition-based maintenance},
  author={Andrew K.S. Jardine and Daming Lin and Dragan Banjevic},
  journal={Mechanical Systems and Signal Processing},
  volume={20},
  number={7},
  pages={1483--1510},
  year={2006},
  publisher={Elsevier},
  doi={10.1016/j.ymssp.2005.09.012}
}

@article{lee2024activity,
  title={Activity-Guided Industrial Anomalous Sound Detection against Interferences},
  author={Lee, Yunjoo and Kim, Jaechang and Ok, Jungseul},
  journal={arXiv preprint arXiv:2409.01885},
  year={2024}
}

@INPROCEEDINGS{hojj2022cont,
  author={Hojjati, Hadi and Armanfard, Narges},
  booktitle={ICASSP 2022 - 2022 IEEE International Conference on Acoustics, Speech and Signal Processing (ICASSP)}, 
  title={Self-Supervised Acoustic Anomaly Detection Via Contrastive Learning}, 
  year={2022},
  volume={},
  number={},
  pages={3253-3257},
  doi={10.1109/ICASSP43922.2022.9746207}}

@article{ANVAR2023103872,
title = {A novel application of deep transfer learning with audio pre-trained models in pump audio fault detection},
journal = {Computers in Industry},
volume = {147},
pages = {103872},
year = {2023},
issn = {0166-3615},
author = {Ali Akbar Taghizadeh Anvar and Hossein Mohammadi},
}

@inproceedings{Deng_2022,
  author={Deng, Jiankang and Guo, Jia and Xue, Niannan and Zafeiriou, Stefanos},
  booktitle={2019 IEEE/CVF Conference on Computer Vision and Pattern Recognition (CVPR)}, 
  title={ArcFace: Additive Angular Margin Loss for Deep Face Recognition}, 
  year={2019},
  volume={},
  number={},
  pages={4685-4694},
  doi={10.1109/CVPR.2019.00482}}

@inproceedings{kolesnikov2019revisiting,
  title={Revisiting self-supervised visual representation learning},
  author={Kolesnikov, Alexander and Zhai, Xiaohua and Beyer, Lucas},
  booktitle={Proceedings of the IEEE/CVF conference on computer vision and pattern recognition},
  pages={1920--1929},
  year={2019}
}

@article{lee2018simple,
  title={A simple unified framework for detecting out-of-distribution samples and adversarial attacks},
  author={Lee, Kimin and Lee, Kibok and Lee, Honglak and Shin, Jinwoo},
  journal={Advances in neural information processing systems},
  volume={31},
  year={2018}
}

@article{rubin2017recognizing,
  title={Recognizing abnormal heart sounds using deep learning},
  author={Rubin, Jonathan and Abreu, Rui and Ganguli, Anurag and Nelaturi, Saigopal and Matei, Ion and Sricharan, Kumar},
  journal={arXiv preprint arXiv:1707.04642},
  year={2017}
}

@ARTICLE{Park2018,
  author={Park, Daehyung and Hoshi, Yuuna and Kemp, Charles C.},
  journal={IEEE Robotics and Automation Letters}, 
  title={A Multimodal Anomaly Detector for Robot-Assisted Feeding Using an LSTM-Based Variational Autoencoder}, 
  year={2018},
  volume={3},
  number={3},
  pages={1544-1551},
  doi={10.1109/LRA.2018.2801475}}

@Article{Lee2016,
AUTHOR = {Lee, Jonguk and Choi, Heesu and Park, Daihee and Chung, Yongwha and Kim, Hee-Young and Yoon, Sukhan},
TITLE = {Fault Detection and Diagnosis of Railway Point Machines by Sound Analysis},
JOURNAL = {Sensors},
VOLUME = {16},
YEAR = {2016},
NUMBER = {4},
ARTICLE-NUMBER = {549},
PubMedID = {27092509},
ISSN = {1424-8220},
DOI = {10.3390/s16040549}
}

@inproceedings{Dohi2022-2,
    author = "Dohi, Kota and Imoto, Keisuke and Harada, Noboru and Niizumi, Daisuke and Koizumi, Yuma and Nishida, Tomoya and Purohit, Harsh and Tanabe, Ryo and Endo, Takashi and Yamamoto, Masaaki and Kawaguchi, Yohei",
    title = "Description and Discussion on {DCASE} 2022 Challenge Task 2: Unsupervised Anomalous Sound Detection for Machine Condition Monitoring Applying Domain Generalization Techniques",
    booktitle = "Proceedings of the 7th Detection and Classification of Acoustic Scenes and Events 2022 Workshop (DCASE2022)",
    address = "Nancy, France",
    month = "November",
    year = "2022",
    pages = "1--5",
}

@article{guan2023transformer,
  title={Transformer-based autoencoder with ID constraint for unsupervised anomalous sound detection},
  author={Guan, Jian and Liu, Youde and Kong, Qiuqiang and Xiao, Feiyang and Zhu, Qiaoxi and Tian, Jiantong and Wang, Wenwu},
  journal={EURASIP journal on audio, speech, and music processing},
  volume={2023},
  number={1},
  pages={42},
  year={2023},
  publisher={Springer}
}

@article{kuroyanagi2021anomalous,
  title={Anomalous sound detection with ensemble of autoencoder and binary classification approaches},
  author={Kuroyanagi, Ibuki and Hayashi, Tomoki and Adachi, Yusuke and Yoshimura, Takenori and Takeda, Kazuya and Toda, Tomoki},
  journal={DCASE2021 Challenge},
  year={2021}
}

@inproceedings{Shin2024,
    author = "Shin, Seunghyeon and Lee, Seokjin",
    title = "Representational Learning for an Anomalous Sound Detection System with Source Separation Model",
    booktitle = "Proceedings of the Detection and Classification of Acoustic Scenes and Events 2024 Workshop (DCASE2024)",
    address = "Tokyo, Japan",
    month = "October",
    year = "2024",
    pages = "146--150",
}

@article{wu2023unsupervised,
  title={Unsupervised anomalous sound detection for industrial monitoring based on ArcFace classifier and gaussian mixture model},
  author={Wu, Ji and Yang, Fei and Hu, Wenkai},
  journal={Applied Acoustics},
  volume={203},
  pages={109188},
  year={2023},
  publisher={Elsevier}
}

@article{wilkinghoff2023angular,
  title={Why do angular margin losses work well for semi-supervised anomalous sound detection?},
  author={Wilkinghoff, Kevin and Kurth, Frank},
  journal={IEEE/ACM Transactions on Audio, Speech, and Language Processing},
  volume={32},
  pages={608--622},
  year={2023},
  publisher={IEEE}
}

@inproceedings{chen2020simple,
  title={A simple framework for contrastive learning of visual representations},
  author={Chen, Ting and Kornblith, Simon and Norouzi, Mohammad and Hinton, Geoffrey},
  booktitle={International conference on machine learning},
  pages={1597--1607},
  year={2020},
  organization={PmLR}
}

@inproceedings{chen2021exploring,
  title={Exploring simple siamese representation learning},
  author={Chen, Xinlei and He, Kaiming},
  booktitle={Proceedings of the IEEE/CVF conference on computer vision and pattern recognition},
  pages={15750--15758},
  year={2021}
}

@inproceedings{guan2023anomalous,
  title={Anomalous sound detection using audio representation with machine id based contrastive learning pretraining},
  author={Guan, Jian and Xiao, Feiyang and Liu, Youde and Zhu, Qiaoxi and Wang, Wenwu},
  booktitle={ICASSP 2023-2023 IEEE International Conference on Acoustics, Speech and Signal Processing (ICASSP)},
  pages={1--5},
  year={2023},
  organization={IEEE}
}

@article{chen2022beats,
  title={Beats: Audio pre-training with acoustic tokenizers},
  author={Chen, Sanyuan and Wu, Yu and Wang, Chengyi and Liu, Shujie and Tompkins, Daniel and Chen, Zhuo and Wei, Furu},
  journal={arXiv preprint arXiv:2212.09058},
  year={2022}
}

@article{chen2024eat,
  title={EAT: Self-supervised pre-training with efficient audio transformer},
  author={Chen, Wenxi and Liang, Yuzhe and Ma, Ziyang and Zheng, Zhisheng and Chen, Xie},
  journal={arXiv preprint arXiv:2401.03497},
  year={2024}
}

@techreport{WangMYPS2025,
    Author = "Wang, Lei",
    title = "PRE-TRAINED MODEL ENHANCED ANOMALOUS SOUND DETECTION SYSTEM FOR DCASE2025 TASK2",
    institution = "DCASE2025 Challenge",
    year = "2025",
    month = "June",
}

@techreport{SaengthongSCITOK2025,
    Author = "Saengthong, Phurich and Shinozaki, Takahiro",
    title = "GENREP FOR FIRST-SHOT UNSUPERVISED ANOMALOUS SOUND DETECTION OF DCASE 2025 CHALLENGE",
    institution = "DCASE2025 Challenge",
    year = "2025",
    month = "June",
}

@article{maaten2008visualizing,
  title={Visualizing data using t-SNE},
  author={Maaten, Laurens van der and Hinton, Geoffrey},
  journal={Journal of machine learning research},
  volume={9},
  number={Nov},
  pages={2579--2605},
  year={2008}
}

@article{mcinnes2018umap,
  title={Umap: Uniform manifold approximation and projection for dimension reduction},
  author={McInnes, Leland and Healy, John and Melville, James},
  journal={arXiv preprint arXiv:1802.03426},
  year={2018}
}

@article{alain2016understanding,
  title={Understanding intermediate layers using linear classifier probes},
  author={Alain, Guillaume and Bengio, Yoshua},
  journal={arXiv preprint arXiv:1610.01644},
  year={2016}
}

@article{hewitt2019designing,
  title={Designing and interpreting probes with control tasks},
  author={Hewitt, John and Liang, Percy},
  journal={arXiv preprint arXiv:1909.03368},
  year={2019}
}

@inproceedings{wang2020understanding,
  title={Understanding contrastive representation learning through alignment and uniformity on the hypersphere},
  author={Wang, Tongzhou and Isola, Phillip},
  booktitle={International conference on machine learning},
  pages={9929--9939},
  year={2020},
  organization={PMLR}
}

@inproceedings{ericsson2021well,
  title={How well do self-supervised models transfer?},
  author={Ericsson, Linus and Gouk, Henry and Hospedales, Timothy M},
  booktitle={Proceedings of the IEEE/CVF conference on computer vision and pattern recognition},
  pages={5414--5423},
  year={2021}
}

@article{loshchilov2017decoupled,
  title={Decoupled weight decay regularization},
  author={Loshchilov, Ilya and Hutter, Frank},
  journal={arXiv preprint arXiv:1711.05101},
  year={2017}
}

@inproceedings{he2016deep,
  title={Deep residual learning for image recognition},
  author={He, Kaiming and Zhang, Xiangyu and Ren, Shaoqing and Sun, Jian},
  booktitle={Proceedings of the IEEE conference on computer vision and pattern recognition},
  pages={770--778},
  year={2016}
}

@article{shimonishi2023anomalous,
  title={Anomalous Sound Detection Based on Sound Separation},
  author={Shimonishi, Kanta and Dohi, Kota and Kawaguchi, Yohei},
  journal={INTERSPEECH 2023},
  year={2023},
  publisher={ISCA}
}

@ARTICLE{CMGAN,
  author={Abdulatif, Sherif and Cao, Ruizhe and Yang, Bin},
  journal={IEEE/ACM Transactions on Audio, Speech, and Language Processing}, 
  title={CMGAN: Conformer-Based Metric-GAN for Monaural Speech Enhancement}, 
  year={2024},
  volume={32},
  number={},
  pages={2477-2493},
  doi={10.1109/TASLP.2024.3393718}}

@inproceedings{gulati2020conformer,
  title={Conformer: Convolution-augmented Transformer for Speech Recognition},
  author={Gulati, Anmol and Qin, James and Chiu, Chung-Cheng and Parmar, Niki and Zhang, Yu and Yu, Jiahui and Han, Wei and Wang, Shibo and Zhang, Zhengdong and Wu, Yonghui and others},
  booktitle={Proc. Interspeech 2020},
  pages={5036--5040},
  year={2020}
}

@inproceedings{gemmeke2017audio,
  title={Audio set: An ontology and human-labeled dataset for audio events},
  author={Gemmeke, Jort F and Ellis, Daniel PW and Freedman, Dylan and Jansen, Aren and Lawrence, Wade and Moore, R Channing and Plakal, Manoj and Ritter, Marvin},
  booktitle={2017 IEEE international conference on acoustics, speech and signal processing (ICASSP)},
  pages={776--780},
  year={2017},
  organization={IEEE}
}

@inproceedings{le2019sdr,
  title={SDR--half-baked or well done?},
  author={Le Roux, Jonathan and Wisdom, Scott and Erdogan, Hakan and Hershey, John R},
  booktitle={ICASSP 2019-2019 IEEE International Conference on Acoustics, Speech and Signal Processing (ICASSP)},
  pages={626--630},
  year={2019},
  organization={IEEE}
}

@misc{tan2020efficientnet,
      title={EfficientNet: Rethinking Model Scaling for Convolutional Neural Networks},
      author={Mingxing Tan and Quoc V. Le},
      year={2020},
      eprint={1905.11946},
      archivePrefix={arXiv},
      primaryClass={cs.LG}
}

@inproceedings{Nishida2025DCASE,
    author = "Nishida, Tomoya and Noboru, Harada and Niizumi, Daisuke and Albertini, Davide and Sannino, Roberto and Pradolini, Simone and Augusti, Filippo and Imoto, Keisuke and Dohi, Kota and Purohit, Harsh and Endo, Takashi and Kawaguchi, Yohei",
    title = "Description and Discussion on DCASE 2025 Challenge Task 2: First-Shot Unsupervised Anomalous Sound Detection for Machine Condition Monitoring",
    booktitle = "Proceedings of the 10th Workshop on Detection and Classification of Acoustic Scenes and Events (DCASE 2025)",
    address = "Barcelona, Spain",
    month = "October",
    year = "2025",
    pages = "55--59",
    isbn = "978-84-09-77652-8",
    doi = "10.5281/zenodo.17251589"
}

@article{Nishida2026DCASE,
  title={Description and Discussion on DCASE 2026 Challenge Task 2: Noise-aware Unsupervised Anomalous Sound Detection for Machine Condition Monitoring},
  author={Nishida, Tomoya and Harada, Noboru and Takeuchi, Daiki and Niizumi, Daisuke and Imoto, Keisuke and Dohi, Kota and Purohit, Harsh and Endo, Takashi and Kawaguchi, Yohei},
  journal={arXiv preprint arXiv:2606.01578},
  year={2026}
}

@article{Fujimura2025ASDKit,
  title={ASDKit: A toolkit for comprehensive evaluation of anomalous sound detection methods},
  author={Fujimura, Takuya and Wilkinghoff, Kevin and Imoto, Keisuke and Toda, Tomoki},
  journal={arXiv preprint arXiv:2507.10264},
  year={2025}
}

@article{Han2025GeneralizedSSLASD,
  title={Exploring self-supervised audio models for generalized anomalous sound detection},
  author={Han, Bing and Jiang, Anbai and Zheng, Xinhu and Zhang, Wei-Qiang and Liu, Jia and Fan, Pingyi and Qian, Yanmin},
  journal={IEEE Transactions on Audio, Speech and Language Processing},
  year={2025},
  publisher={IEEE}
}

@article{Wilkinghoff2026TemporalPooling,
  title={Temporal pooling strategies for training-free anomalous sound detection with self-supervised audio embeddings},
  author={Wilkinghoff, Kevin and Yadav, Sarthak and Tan, Zheng-Hua},
  journal={arXiv preprint arXiv:2603.04605},
  year={2026}
}

@inproceedings{Dinkel2024CED,
  title={CED: Consistent ensemble distillation for audio tagging},
  author={Dinkel, Heinrich and Wang, Yongqing and Yan, Zhiyong and Zhang, Junbo and Wang, Yujun},
  booktitle={ICASSP 2024-2024 IEEE International Conference on Acoustics, Speech and Signal Processing (ICASSP)},
  pages={291--295},
  year={2024},
  organization={IEEE}
}

@inproceedings{Jiang2025GLAMASD,
  title={An Effective Anomalous Sound Detection Method Based on Global and Local Attribute Mining},
  author={Jiang, Nan and Song, Yan and Gu, Qing and Song, Haoyu and Dai, Lirong and McLoughlin, Ian},
  booktitle={Proc. Interspeech 2025},
  pages={2620--2624},
  year={2025}
}

@Article{heartbeat,
AUTHOR = {Guerra, Roger de T. and Yamaguchi, Cristina K. and Stefenon, Stefano F. and Coelho, Leandro dos S. and Mariani, Viviana C.},
TITLE = {Deep Learning Approach for Automatic Heartbeat Classification},
JOURNAL = {Sensors},
VOLUME = {25},
YEAR = {2025},
NUMBER = {5},
ARTICLE-NUMBER = {1400},
URL = {https://www.mdpi.com/1424-8220/25/5/1400},
PubMedID = {40096255},
ISSN = {1424-8220},
DOI = {10.3390/s25051400}
}



\end{document}